\documentclass[prd,twocolumn,showpacs,amsmath,amssymb]{revtex4-1}
\usepackage{graphicx}
\usepackage{dcolumn}
\usepackage{bm}
\usepackage{comment}
\usepackage{multirow}
\usepackage{ulem}
\usepackage[colorlinks,
linkcolor=blue,
anchorcolor=blue,
citecolor=blue
]{hyperref}
\usepackage{rotating}
\usepackage{color}
\usepackage{xcolor}

\begin{document}
\normalsize
\parskip=5pt plus 1pt minus 1pt
\title{Cross sections for the reactions $e^+e^-\rightarrow K^+K^-\pi^+\pi^-(\pi^0)$, $K^+K^-K^+K^-(\pi^0)$, $\pi^+\pi^-\pi^+\pi^-(\pi^0)$, $p\bar{p}\pi^+\pi^-(\pi^0)$ in the energy region between 3.773 and 4.600 GeV}
\author{\small M.~Ablikim$^{1}$, M.~N.~Achasov$^{10,c}$, P.~Adlarson$^{67}$, S. ~Ahmed$^{15}$, M.~Albrecht$^{4}$, R.~Aliberti$^{28}$, A.~Amoroso$^{66A,66C}$, M.~R.~An$^{32}$, Q.~An$^{63,49}$, X.~H.~Bai$^{57}$
% ...
\\
\vspace{0.2cm}
(BESIII Collaboration)\\
\vspace{0.2cm} {\it
$^{1}$ Institute of High Energy Physics, Beijing 100049, People's Republic of China\\
% ...
$^{4}$ Bochum Ruhr-University, D-44780 Bochum, Germany\\
% ...
$^{10}$ G.I. Budker Institute of Nuclear Physics SB RAS (BINP), Novosibirsk 630090, Russia\\
% ...
$^{15}$ Helmholtz Institute Mainz, Staudinger Weg 18, D-55099 Mainz, Germany\\
% ...
$^{28}$ Johannes Gutenberg University of Mainz, Johann-Joachim-Becher-Weg 45, D-55099 Mainz, Germany\\
% ...
$^{32}$ Liaoning Normal University, Dalian 116029, People's Republic of China\\
% ...
$^{49}$ State Key Laboratory of Particle Detection and Electronics, Beijing 100049, Hefei 230026, People's Republic of China\\
% ...
$^{57}$ University of Jinan, Jinan 250022, People's Republic of China\\
% ...
$^{63}$ University of Science and Technology of China, Hefei 230026, People's Republic of China\\
% ...
$^{66}$ University of Turin and INFN, (A)University of Turin, I-10125, Turin, Italy; (B)University of Eastern Piedmont, I-15121, Alessandria, Italy; (C)INFN, I-10125, Turin, Italy\\
$^{67}$ Uppsala University, Box 516, SE-75120 Uppsala, Sweden\\
% ...
$^{c}$ Also at the Novosibirsk State University, Novosibirsk, 630090, Russia\\
% ...
}
}
\date{\today}
\begin{abstract}
Using the data samples collected in the energy range from 3.773 to 4.600 GeV with the
BESIII detector at the BEPCII collider, we measure the dressed cross sections as a function of center-of-mass energy  
for $e^+e^-\rightarrow K^+K^-\pi^+\pi^-(\pi^0)$, $K^+K^-K^+K^-(\pi^0)$, 
	$\pi^+\pi^-\pi^+\pi^-(\pi^0)$, and $p\bar{p}\pi^+\pi^-(\pi^0)$. 
The cross sections for $e^+e^-\rightarrow K^+K^-K^+K^-\pi^0$,
 $p\bar{p}\pi^+\pi^-(\pi^0)$ are the first measurements. Cross sections for the other five channels are much
more precise than previous results in this energy region. 
We also search for charmonium and charmonium-like resonances, such as the $Y(4230)$, decaying into the same final states.
We find evidence of the $\psi(4040)$ decaying to $\pi^+\pi^-\pi^+\pi^-\pi^0$ with a statistical significance of $3.6\sigma$. 
Upper limits are provided for other decays since no clear signals are observed.
\end{abstract}
\maketitle
\section{Introduction}
The energy region above open-charm threshold provides a place to test and develop quantum chromodynamics (QCD). In the past decade, a series of charmonium-like states~\cite{XYZ} were observed, such as the $Y(4260)$ state discovered by the $BABAR$ Collaboration through the initial-state radiation (ISR) process $\gamma_{\rm ISR}\pi^+\pi^-J/\psi$~\cite{Y4260BaBar}, and confirmed by CLEO~\cite{Y4260CLEO}, Belle~\cite{Y4260Belle} and BESIII~\cite{Y4260BESIII} in the same process. 
A recent precise measurement of $e^+e^-\rightarrow\pi^{+}\pi^{-}J/\psi$~\cite{Y4220_Jpsipipi} shows that the $Y(4260)$ consists of two resonances; the narrower resonance at lower mass is called $Y(4230)$. The $Y(4230)$ has also been reported by BESIII in the study of the $e^+e^-\rightarrow\omega\chi_{c0}$~\cite{Y4220_omegachic}, $\pi^+\pi^{-}h_{c}$~\cite{Y4220_hcpipi}, and $\pi^+D^0D^{*-}$~\cite{Y4220_DDstarpi} cross sections.\par 
Contrary to the conventional charmonium states, the $Y(4230)$
 strongly couples to $\pi\pi J/\psi$~\cite{Y4220_Jpsipipi,Y4220_DDstarpi}. 
This is also true for other $Y$ states, e.g. $Y(4360)$, $Y(4660)$.
The discoveries of those exotic particles have prompted further
investigations of the center-of-mass (c.m.) energy-dependent cross sections~\cite{ref7}. To explore the
nature of these exotic particles, a variety of decay modes have been studied, such as open-charm processes ($D^{(*)}\bar{D}^{(*)}$\cite{ref013,ref014,ref017,ref018}, $D^{(*)}\bar{D}^{(*)}\pi$\cite{ref017,ref016}, $D\bar{D_2}^{*}(2460)$\cite{ref015} and $D_s^{(*)+}D_s^{(*)-}$\cite{ref017,ref019}), 
and transitions to other charmonium states ($\pi^+\pi^- J/\psi$\cite{ref_pipiJpsi}, $\pi^0\pi^0 J/\psi$\cite{ref_pi0pi0Jpsi}, $\eta J/\psi$\cite{ref021}, $\pi^+\pi^-h_c$\cite{ref_pipihc}, $\pi^+\pi^-\psi(3686$)\cite{ref020} and $\omega\chi_{c0}$\cite{ref_omegachic}). 
Many light hadron final states ($K^+K^-\pi^+\pi^-$\cite{ref_old,ref_new}, $K^+K^-K^+K^-$\cite{ref_old,ref_new,ref_besiii}, $\pi^+\pi^-\pi^+\pi^-$\cite{ref_old}, $\phi f_{0}(980)$\cite{ref022}, $K^+K^-\eta$\cite{ref023}, $K^+K^-\pi^0$\cite{ref023}, $K_s^0K^\pm\pi^\mp$\cite{ref023}, $p\bar{p}$\cite{ref024}, etc.) 
have also been studied. However, no light hadronic decays of the $Y$ states or conventional charmonium resonances above 4~GeV have yet been observed\cite{ref_old,ref_new,ref_besiii,ref022,ref023,ref024}.
The continued search for light hadron decays may help clarify the nature of exotic states and charmonium resonances\cite{reftheory1,reftheory2}.\par
To study charmonium and charmonium-like particles, the BESIII detector has collected the world's largest data samples in the energy region between 3.773 GeV and 4.600~GeV. Based on those data sets, we analyze the dressed cross sections for the processes $e^+e^-\rightarrow K^+K^-\pi^+\pi^-(\pi^0)$, $K^+K^-K^+K^-(\pi^0)$, $\pi^+\pi^-\pi^+\pi^-(\pi^0)$, and $p\bar{p}\pi^+\pi^-(\pi^0)$ and search for possible structures, such as charmonium or $Y$ states, in the line shapes of the $e^{+}e^{-}$ cross sections.\par
This paper is organized as follows. In Sec.~\ref{sec:detector}, we describe the BESIII detector. The data and Monte Carlo (MC) samples are introduced in Sec.~\ref{sec:data}. In Sec.~\ref{sec:Nobs}, we describe the requirements for the selection of signal events. In Sec.~\ref{sec:Nbkg}, we present the measurement of the number of background events. The determination of $\epsilon^{0}$ and $\kappa$ are described in Sec.~\ref{sec:Eff} and Sec.~\ref{sec:kappa}, respectively.  Sec.~\ref{sec:sys} discusses systematics uncertainties and a summary is presented in Sec.~\ref{sec:summary}.\par
\section{Detector}
\label{sec:detector}
The BESIII detector is a magnetic
spectrometer~\cite{Ablikim:2009aa} located at the Beijing Electron
Positron Collider (BEPCII)~\cite{Yu:IPAC2016-TUYA01}. The
cylindrical core of the BESIII detector consists of a helium-based
 multilayer drift chamber (MDC), a plastic scintillator time-of-flight
system (TOF), and a CsI(Tl) electromagnetic calorimeter (EMC),
which are all enclosed in a superconducting solenoidal magnet
providing a 1.0~T magnetic field. The solenoid is supported by an
octagonal flux-return yoke with resistive plate counter muon
identifier modules interleaved with steel. The acceptance of
charged particles and photons is 93\% over $4\pi$ solid angle. The
charged-particle momentum resolution at $1~{\rm GeV}/c$ is
$0.5\%$, and the $dE/dx$ resolution is $6\%$ for the electrons
from Bhabha scattering. The EMC measures photon energies with a
resolution of $2.5\%$ ($5\%$) at $1$~GeV in the barrel (end cap)
region. The time resolution of the TOF barrel part is 68~ps, while
that of the end cap part is 110~ps. The end cap 
TOF system is upgraded in 2015 with multi-gap resistive plate 
chamber technology, providing a time resolution of 60 ps\cite{etof}.  
This improvement affects $57\%$ of the data used in this work.\par
\section{Data and Monte Carlo samples}
\label{sec:data}
In this work, we analyzed the data sets taken at c.m.~energies 
from 3.773 to 4.600 GeV. Measurements of c.m.~energies and 
luminosities are described elsewhere~\cite{CMxyz,Lumxyz}.\par 
The response of the BESIII detector is modeled with
MC simulations using the software frame work BOOST~\cite{REF25}
based on GEANT4~\cite{REF26}, which includes the geometry and material 
description of the BESIII detectors, the detector response and 
digitization models, as well as a database that keeps track of 
the running conditions and the detector performance.\par
The signal MC samples at all c.m.~energies 
are generated with a phase space (PHSP) model.
The inclusive MC samples generated at different c.m.~energies 
are used to study the potential backgrounds.
The inclusive MC samples consist of the production
of open charm processes, the ISR production of vector
charmonium states, and the continuum processes
incorporated in KKMC~\cite{REF27}. The known decay modes are
modeled with EvtGen~\cite{REF28} using branching fractions taken 
from the Particle Data Group (PDG)~\cite{PDG},  the
remaining unknown decays from the charmonium states
with {\sc lundcharm}~\cite{REF30}, and the cross sections for 
the open charm final states are cited from~\cite{REF32,REF33,REF34,REF35,REF36}.   
The FSR from charged final state
particles are incorporated with the {\sc photos} package~\cite{REF31}.
As described in Sec.~\ref{sec:Nbkg}, some ``peaking backgrounds'' could pass our
selection requirements. MC samples for those processes are generated for the 
study of distributions and mis-identification rates.\par
\section{Measurements of cross sections}
\label{sec:method}
For a given c.m.~energy $E_{{\rm cm}}$, the dressed cross sections for $e^+e^-\rightarrow K^+K^-\pi^+\pi^-(\pi^0)$, $K^+K^-K^+K^-(\pi^0)$, $\pi^+\pi^-\pi^+\pi^-(\pi^0)$, and $p\bar{p}\pi^+\pi^-(\pi^0)$ are given by
\begin{equation}
\sigma=\frac{N^{\rm obs}-N^{\rm bkg}}{L \epsilon^{0} \kappa},
\label{equation:cs}
\end{equation}
where $N^{\rm obs}$ is the number of candidate signal 
events observed in the data sample, $N^{\rm bkg}$ is the number of background events, $L$ is the integrated 
luminosity of the data collected at $E_{{\rm cm}}$, $\epsilon^{0}$
is the reconstruction efficiency without considering ISR, and $\kappa$ is the correction factor describing the effect of ISR~\cite{ISR}.\par
\subsection{Selection of Signal Events}
\label{sec:Nobs}
For each channel, all final state particles are reconstructed. To
ensure each track originates from the $e^+e^-$ collision point, the tracks must satisfy $V_r<1.0$~cm
and $|V_z|<10.0$~cm. Here $V_r$ is the distance between the charged track and the beam axis
in the $r-\varphi$ plane, and $|V_z|$ is the coordinate of the charged particle production point along
the beam axis. The polar angles of charged tracks are required to satisfy $|\rm cos~\theta|<0.93$. 
Exactly four good tracks satisfying these criteria are required.\par
For the final states with a $\pi^{0}$, we reconstruct $\pi^{0}$ candidates through $\pi^{0}\rightarrow\gamma\gamma$. Showers must have energy greater than 25 MeV in the barrel region ($|\cos\theta|<0.80$) of the EMC and greater than 50 MeV in the endcaps ($0.86<|\cos\theta|<0.92$). Showers must have timing within 700~ns of the event start time.
For $K^+K^-\pi^+\pi^-$ and $K^+K^-\pi^+\pi^-\pi^0$ final states, we apply particle identification based on $dE/dx$ and TOF measurements to reduce 
multiple combinations of particle hypotheses within candidate events.
We require the kaon candidates have a higher probability to be kaons than pions. Only pions and kaons hypotheses are compared because these are the most serious sources of misidentification.\par
A kinematic fit is applied to candidate events. For signal channels without (with) a $\pi^{0}$, we perform a four-constraint (five-constraint) kinematic fit to the known initial four-momentum (and $\pi^{0}$ mass). 
We require that the $\chi^{2}$ of the kinematic fit is less than 50. If more than one combination per mode satisfies the above selection requirements, only the one with the least $\chi^{2}$ is kept.\par
Background events from the two-photon processes $e^{+}e^{-}\rightarrow e^{+}e^{-}+{\rm hadrons}$, together with $e^{+}e^{-}\rightarrow(\gamma)e^{+}e^{-}$ are rejected using the ratio $E_{\rm EMC}/p$ for each charged track, where $E_{\rm EMC}$ is the energy deposited in the EMC and $p$ is the momentum of the charged track. 
For candidates events from the processes $e^{+}e^{-}\rightarrow K^{+}K^{-}\pi^{+}\pi^{-}, K^{+}K^{-}K^{+}K^{-}, \pi^{+}\pi^{-}\pi^{+}\pi^{-}$ and $\pi^{+}\pi^{-}\pi^{+}\pi^{-}\pi^{0}$, it is required that each track have an $E_{\rm EMC}/p$ less than 0.8. The gamma conversion backgrounds are rejected by applying cuts to $\theta_{\pi^{+}\pi^{-}}$, which is the opening angle between all $\pi^{+}\pi^{-}$ pairs. For candidates events of $e^{+}e^{-}\rightarrow K^{+}K^{-}\pi^{+}\pi^{-}, \pi^{+}\pi^{-}\pi^{+}\pi^{-}$ and $p\bar{p}\pi^{+}\pi^{-}$, it is required that $\cos\theta_{\pi^{+}\pi^{-}}<0.9$. To reduce systematic uncertainty, the distributions of $E_{\rm EMC}/p$ and $\cos\theta_{\pi^{+}\pi^{-}}$ from MC samples have been corrected according to data. Here the MC sample is the PHSP MC reweighted with the amplitude analysis results, which would be described in the Sec.\ref{sec:Eff}. The correction factor is determined using a control sample of the signal process. The distributions of $E_{\rm EMC}/p$ and $\theta_{\pi^{+}\pi^{-}}$ from all the other channels are checked as well but no obvious backgrounds are found.\par
We next check for backgrounds in two-body and three-body invariant mass distributions within each final state. There are obvious backgrounds from $J/\psi$, $\psi(2S)$, $\chi_{c}$, $D$ and $K_{s}^{0}$ decays. Those backgrounds are removed with the requirements summarized in Table~\ref{tab:cuts}.\par 
\begin{table}[htbp]
\centering
\caption{Summary of the cuts applied to the two-body and three-body invariant mass distributions.}
\resizebox{0.48\textwidth}{!}{
\begin{tabular}{|c|c|} \hline
Final state & Cut \\
\hline                                           
$e^+e^-\rightarrow K^+K^-\pi^+\pi^-$         & $|M_{\pi^+\pi^-}-M_{J/\psi}|>15~{\rm MeV}/c^{2}$ \\
                                             & $|M_{K^{\pm}\pi^{\mp}}-M_{D}|>20~{\rm MeV}/c^{2}$ \\
\hline                                           
$e^+e^-\rightarrow\pi^+\pi^-\pi^+\pi^-$      & $|M_{\pi^+\pi^-}-M_{J/\psi}|>15~{\rm MeV}/c^{2}$ \\
                                             & $|M_{\pi^+\pi^-}-M_{\psi(2S)}|>20~{\rm MeV}/c^{2}$ \\
\hline                                           
$e^+e^-\rightarrow p\bar{p}\pi^+\pi^-$       & $|M_{p\bar{p}}-M_{J/\psi}|>15~{\rm MeV}/c^{2}$ \\
                                             & $|M_{p\bar{p}}-M_{\psi(2S)}|>20~{\rm MeV}/c^{2}$ \\
\hline                                           
$e^+e^-\rightarrow K^+K^-\pi^+\pi^-\pi^0$    & $|M_{K^+K^-}-M_{\chi_{c0}}|>50~{\rm MeV}/c^{2}$ \\
                                             & $|M_{K^{\pm}\pi^{\mp}}-M_{D}|>20~{\rm MeV}/c^{2}$ \\
                                             & $|M_{\pi^+\pi^-}-M_{K_{s}^{0}}|>30~{\rm MeV}/c^{2}$ \\
\hline                                           
$e^+e^-\rightarrow\pi^+\pi^-\pi^+\pi^-\pi^0$ & $|M_{\pi^+\pi^-}-M_{J/\psi}|>15~{\rm MeV}/c^{2}$ \\
                                             & $|M_{\pi^+\pi^-}-M_{\chi_{c0}}|>50~{\rm MeV}/c^{2}$ \\
                                             & $|M_{\pi^+\pi^-}-M_{K_{s}^{0}}|>30~{\rm MeV}/c^{2}$ \\
                                             & $|M_{\pi^+\pi^-\pi^0}-M_{J/\psi}|>15~{\rm MeV}/c^{2}$ \\
\hline
\end{tabular}
}
\label{tab:cuts}
\end{table}
\subsection{Background Study}
\label{sec:Nbkg}
Our selected candidate events include both signal events as well as misidentified background events 
from other processes. Potential background  sources include 
$e^+e^-\rightarrow (\gamma)e^+e^-$, $e^+e^-\rightarrow (\gamma)\mu^+\mu^-$, $e^+e^-\rightarrow (\gamma)\tau^+\tau^-$, $e^+e^-\rightarrow\rm{hadrons}$. To study such backgrounds, we analyzed the inclusive MC generated at 4.226 GeV using the same selection requirements for data.  
Such studies show that there are two types of background. The first is from $D$ and $J/\psi$ decays, where the final states are the same as the signal channel. For example, $e^+e^-\rightarrow D^{0}\bar{D}^{0},D^{0}\rightarrow K^{-}\pi^{+},\bar{D}^{0}\rightarrow K^{+}\pi^{-}$ is background process for $e^+e^-\rightarrow K^{+}K^{-}\pi^{+}\pi^{-}$. Although we have applied requirements to veto that background, there are still residual events left. The distributions of the $\chi^{2}$ from the kinematic fit are almost the same as those for the signal events; we call this the ``peaking background.''
The second type of background is from processes where the final states are different from the signal channel.
In this case, the $\chi^{2}$ distributions of these processes are also different than signal; we refer to this background as ``non-peaking background.'' 
To estimate the number of non-peaking background events, a fit is performed to the $\chi^{2}$ from the kinematic fit.\par 
Figure~\ref{fig:FitChi2_Example} shows the fit to the $\chi^{2}$ at 4.226 GeV. The signal shape is from signal MC simulation. The distribution of peaking background is from peaking background MC, the number of which is fixed according to previously measured cross sections{\cite{bkgref,ISRDD} and reconstruction efficiency determined with MC simulation.  The contamination rates from peaking backgrounds are less than $0.8\%$ for all signal final states.
The shape of the non-peaking background is obtained from inclusive background MC samples, and the number of non-peaking background is allowed to float. From this fit, the number of non-peaking background events in the signal region ($\chi^{2}<50$) is obtained.\par 
\subsection{Reconstruction Efficiency}
\label{sec:Eff}
In the two body and three body invariant mass distributions, there are many intermediate states, such as $\rho$, $\omega$, $\phi, \eta, a, b, f, K^*, N, \Lambda, \Delta$. 
Final states with these various resonant intermediate states are also signal processes. To determine the reconstruction efficiency more accurately, we need to consider the relative contributions from those processes.
Using the AmpTools\cite{AmpTools,AmpTools2} package, amplitude analyses are performed to some of the large data samples (at 3.773 GeV, 4.008 GeV, 4.226 GeV, 4.258 GeV, 4.358 GeV, 4.416 GeV and 4.600 GeV), together with MC samples generated at the same energy points.
Relative amplitudes of different intermediate states yielding each final state are determined from these analyses. 
We use the ratios obtained from 4.226 GeV in the determination of reconstruction efficiencies 
for all the energy points. 
Differences in these amplitude ratios are considered as a source of systematic error, 
based on possible small variations from other data sets with large statistics.\par 
Weights are assigned to PHSP signal MC event-by-event according to ratios of squared amplitudes.  
Applying the same selection requirements as used in data analysis, the MC efficiencies $\epsilon^{0}$ are determined from signal MC samples without ISR. 

\subsection{ISR Corrections}
\label{sec:kappa}
The ISR effect is considered using the factor $\kappa$, which is defined as
\begin{equation}
\kappa=\frac{1}{\sigma^{0}\epsilon^{0}}\int\sigma(s') \, \epsilon(x) \, W(x,s) \, dx,
\label{equation:kappa2}
\end{equation}
where $s=E_{\rm cm}^{2}$, $x$ is the radiative photon energy fraction, $s'=s(1-x)$, 
$W(x,s)$ is the radiator function~\cite{ISR} and $\sigma^0$ is the dressed cross section corresponding to $x=0$.\par
Applying the same selection requirements as those applied to the data, $\epsilon(x)$ is determined by analyzing signal MC samples with ISR. The distribution of $\epsilon(x)/\epsilon^0$ for every channel at 4.226 GeV, together with the fit to error function is shown in Fig.~\ref{fig:epsilon_x}. Inserting $\epsilon(x)$ into Eq.~\ref{equation:kappa2}, we calculated the factor $\kappa$.
\begin{figure}[htbp]
  \centering
  \includegraphics[width=0.23\textwidth]{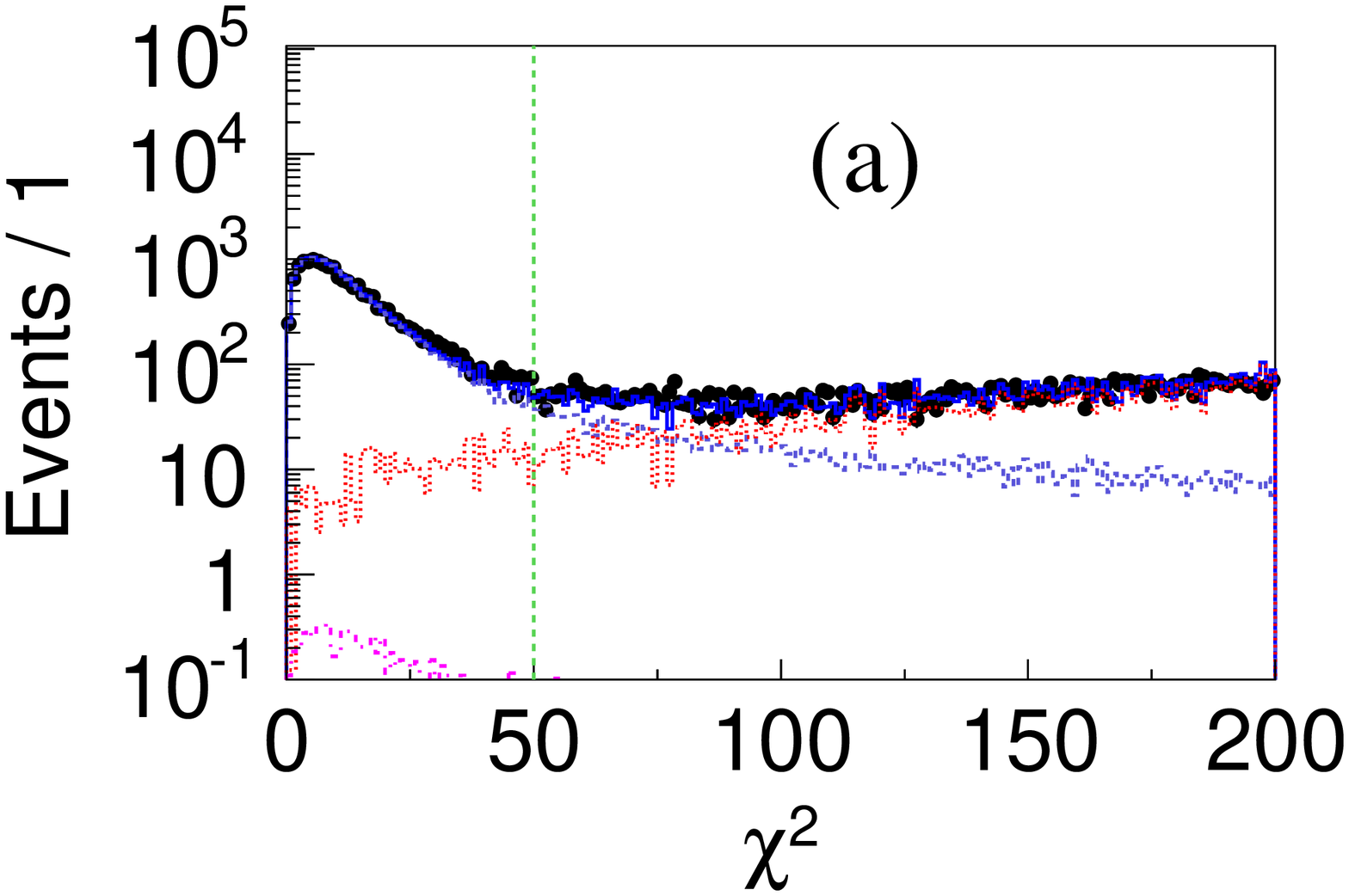}
  \includegraphics[width=0.23\textwidth]{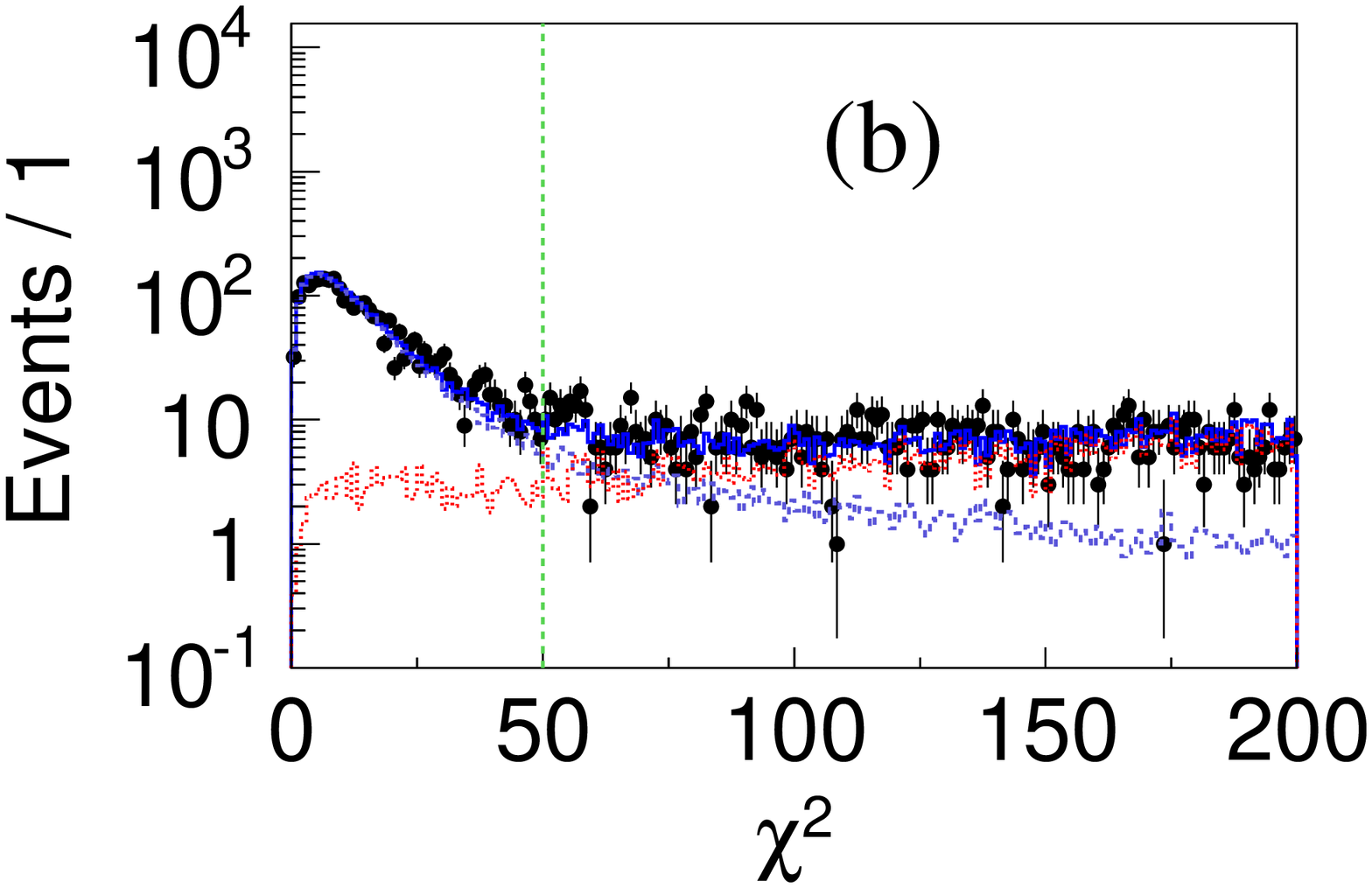}
  \includegraphics[width=0.23\textwidth]{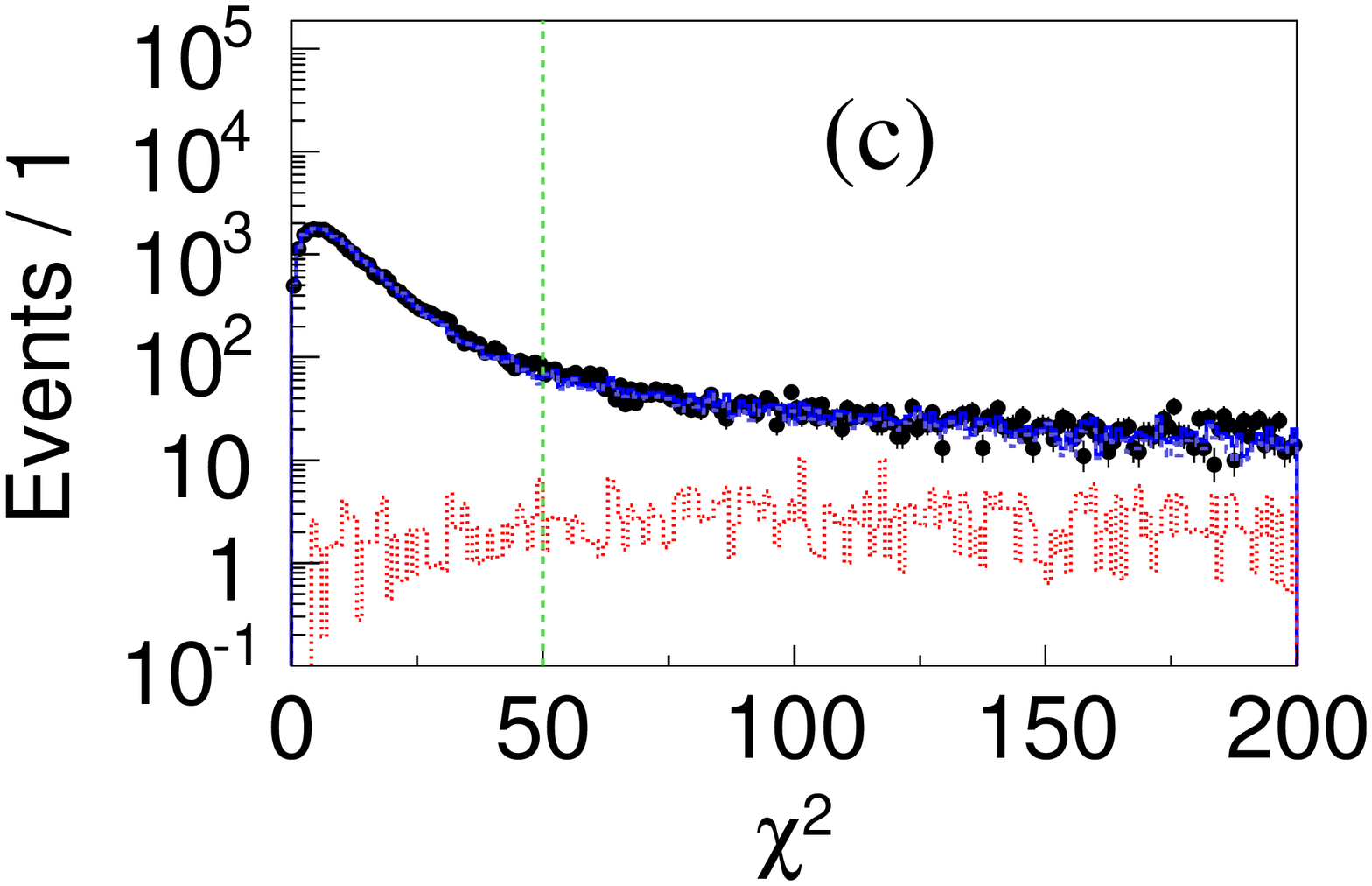}
  \includegraphics[width=0.23\textwidth]{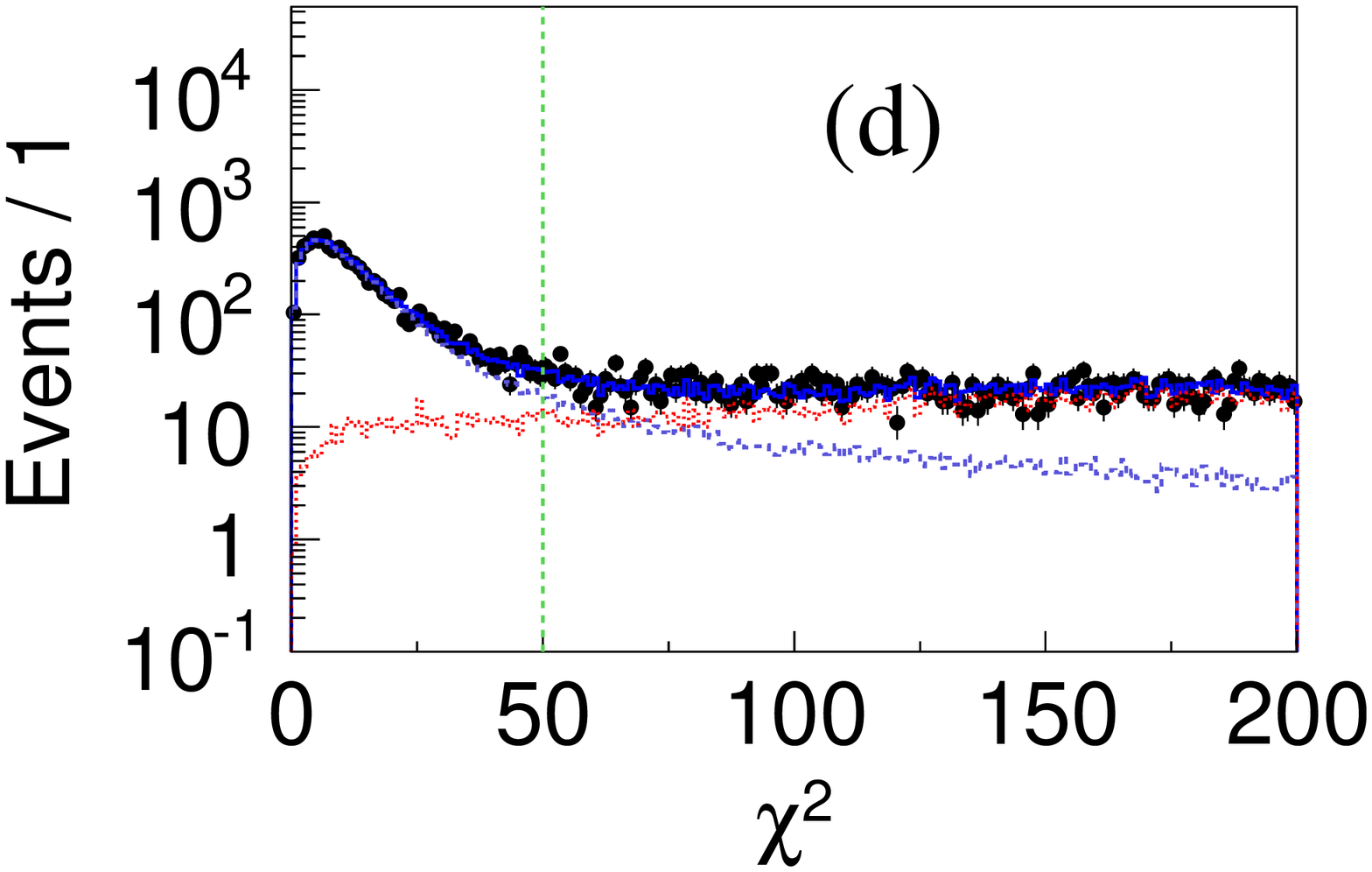}
  \includegraphics[width=0.23\textwidth]{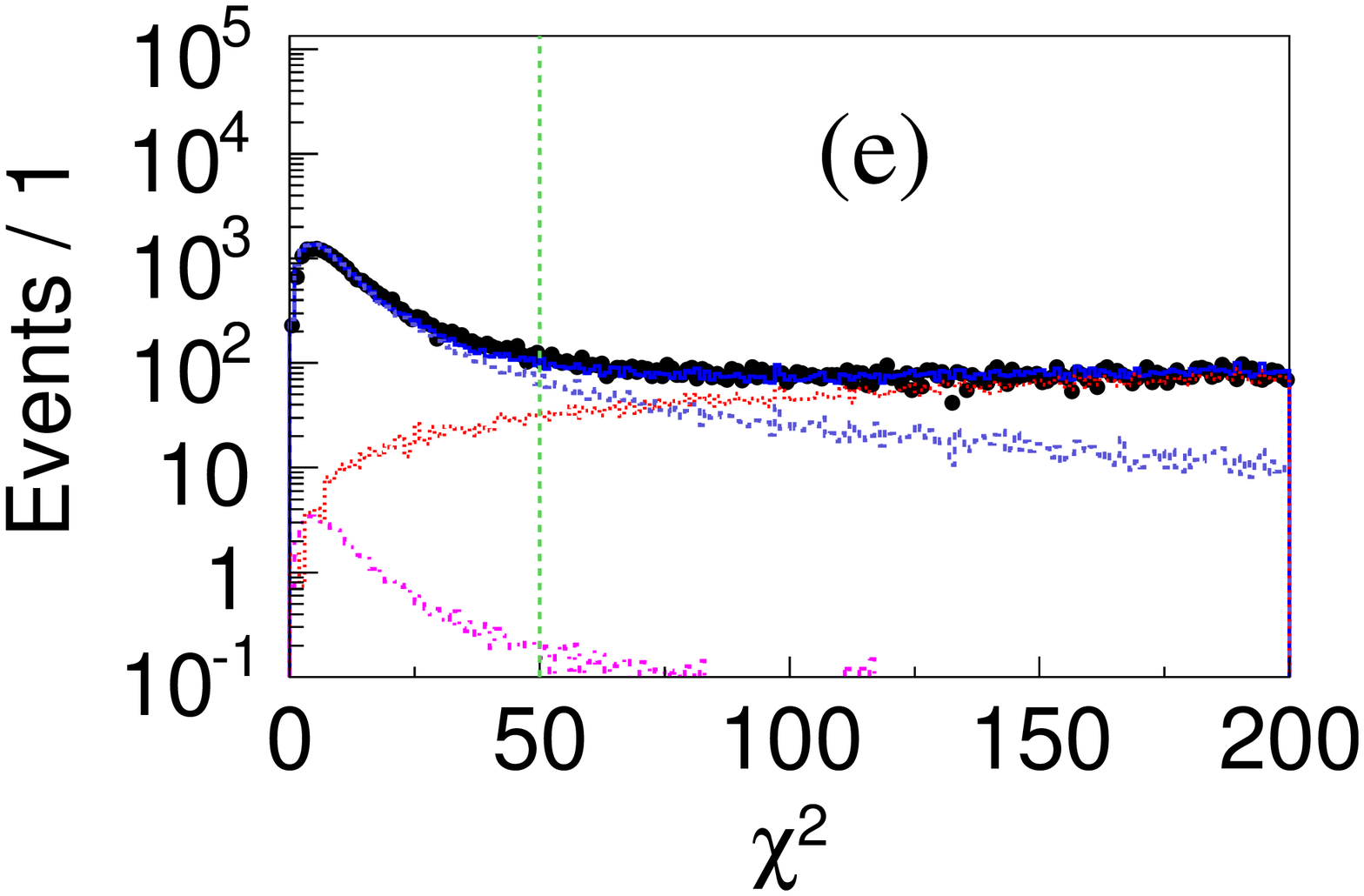}
  \includegraphics[width=0.23\textwidth]{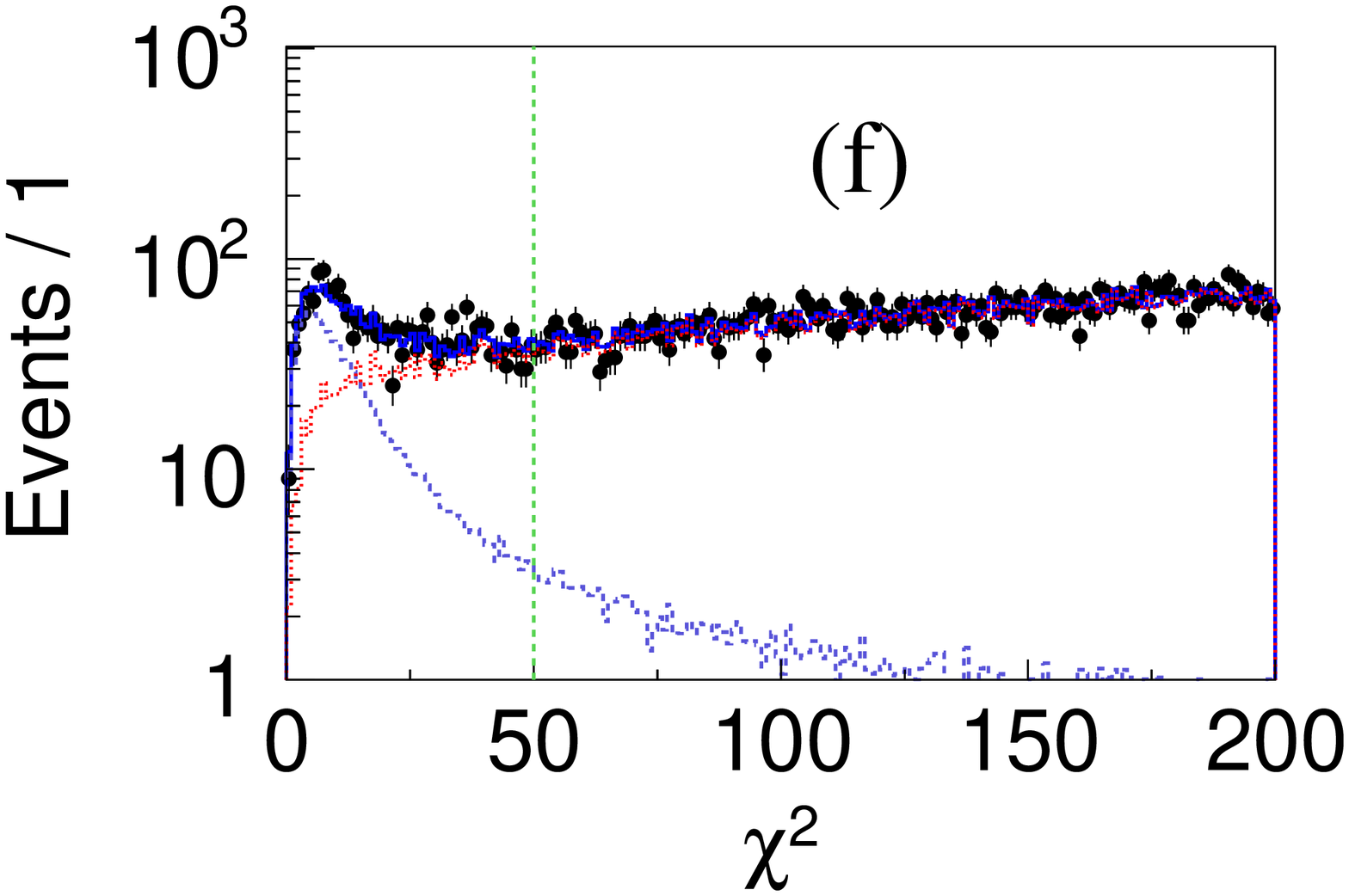}
  \includegraphics[width=0.23\textwidth]{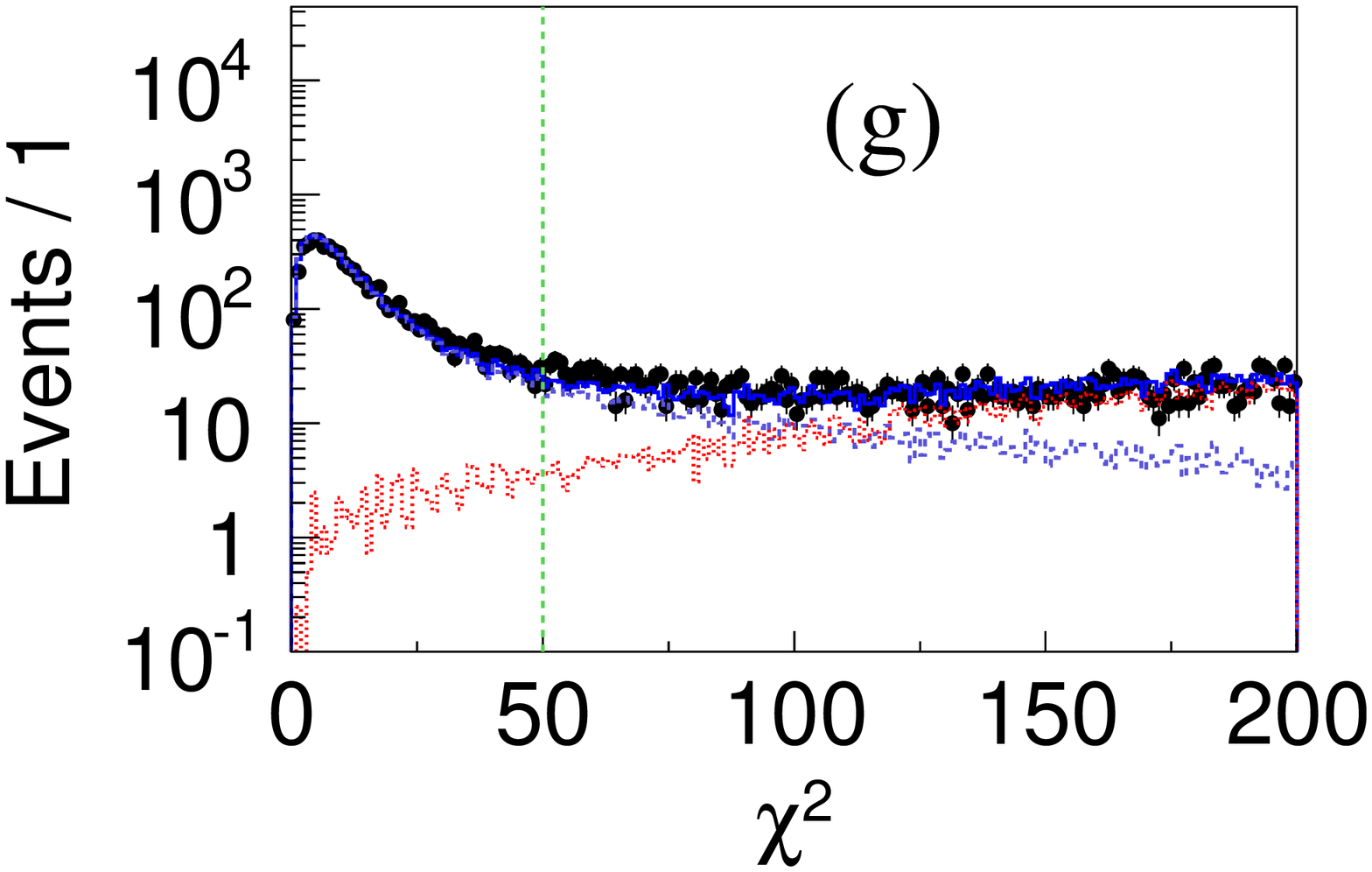}
  \includegraphics[width=0.23\textwidth]{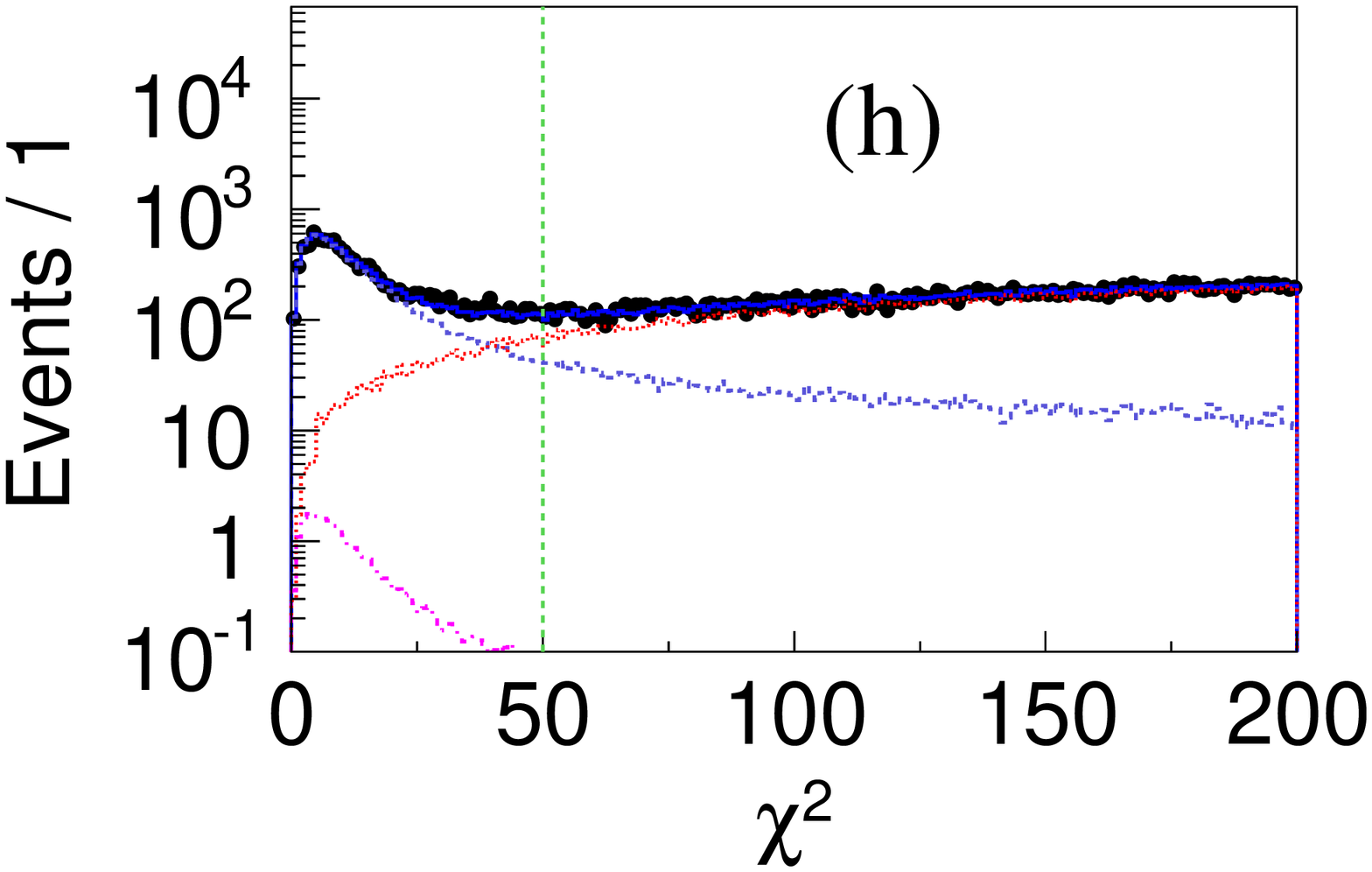}
  \caption{\sf
The fit to the $\chi^{2}$ from the kinematic fit for (a) $e^+e^-\rightarrow K^+K^-\pi^+\pi^-$, (b) $e^+e^-\rightarrow K^+K^-K^+K^-$, (c) $e^+e^-\rightarrow\pi^+\pi^-\pi^+\pi^-$, (d) $e^+e^-\rightarrow p\bar{p}\pi^+\pi^-$, (e) $e^+e^-\rightarrow K^+K^-\pi^+\pi^-\pi^0$, (f) $e^+e^-\rightarrow K^+K^-K^+K^-\pi^0$, (g) $e^+e^-\rightarrow\pi^+\pi^-\pi^+\pi^-\pi^0$, and (h) $e^+e^-\rightarrow p\bar{p}\pi^+\pi^-\pi^0$ at 4.226 GeV. To evaluate the goodness of those fits, we check the $\chi^{2}/ndf$ of those plots, which are 1.75, 1.16, 1.54, 1.18, 1.59, 1.03, 1.36, and 1.50, respectively. 
The number of background in $\chi^2$ signal region, which would be subtracted when calculating the cross section, is obtained from this fit.
The blue dashed line shows the distribution from signal MC. The pink dash-dotted lines and red dotted lines show the distribution from peaking and non-peaking backgrounds, respectively. The green dashed line shows the signal region.
}
  \label{fig:FitChi2_Example}
\end{figure}
\begin{figure}[htbp]
  \centering
  \includegraphics[width=0.23\textwidth]{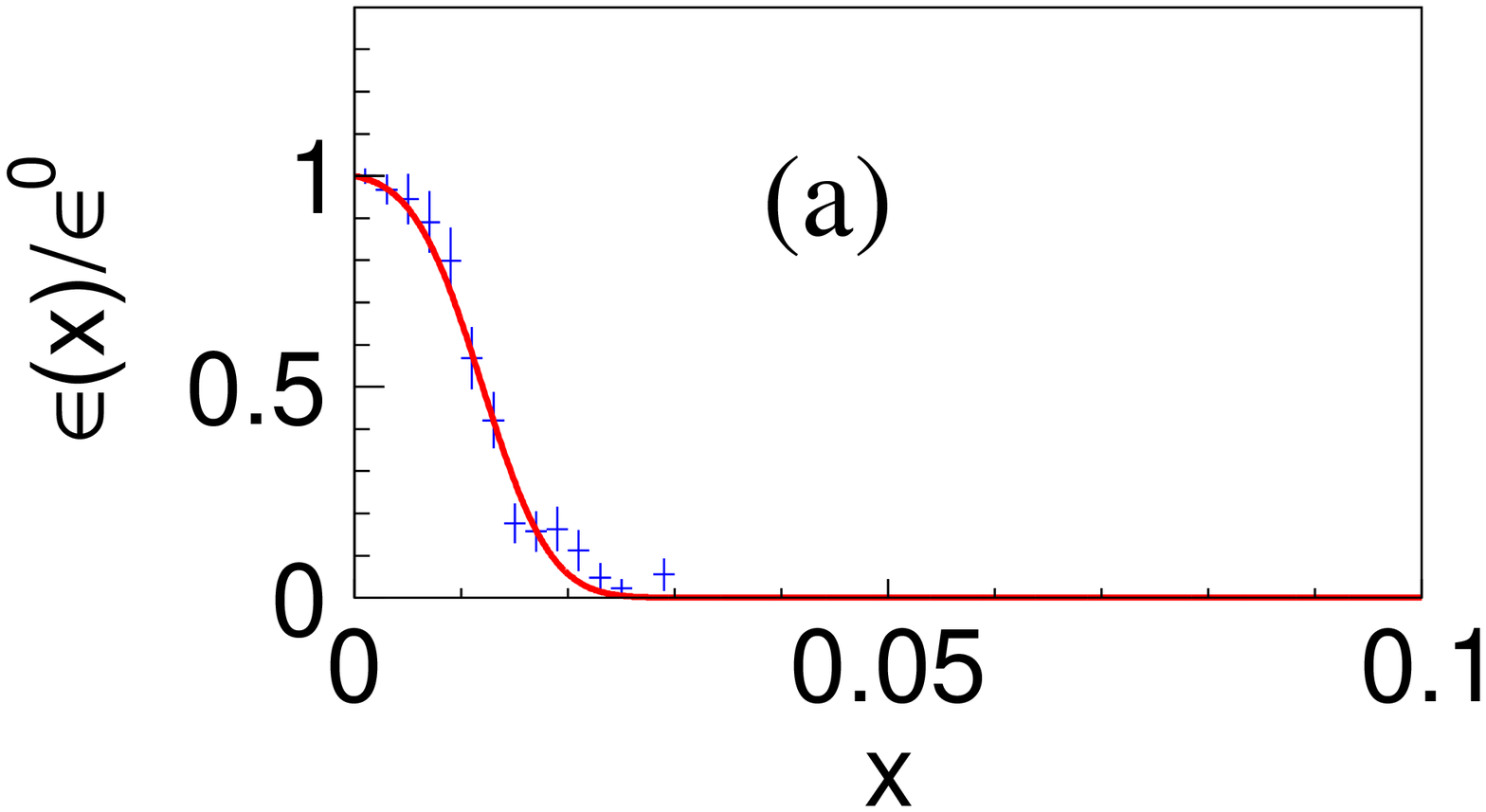} 
  \includegraphics[width=0.23\textwidth]{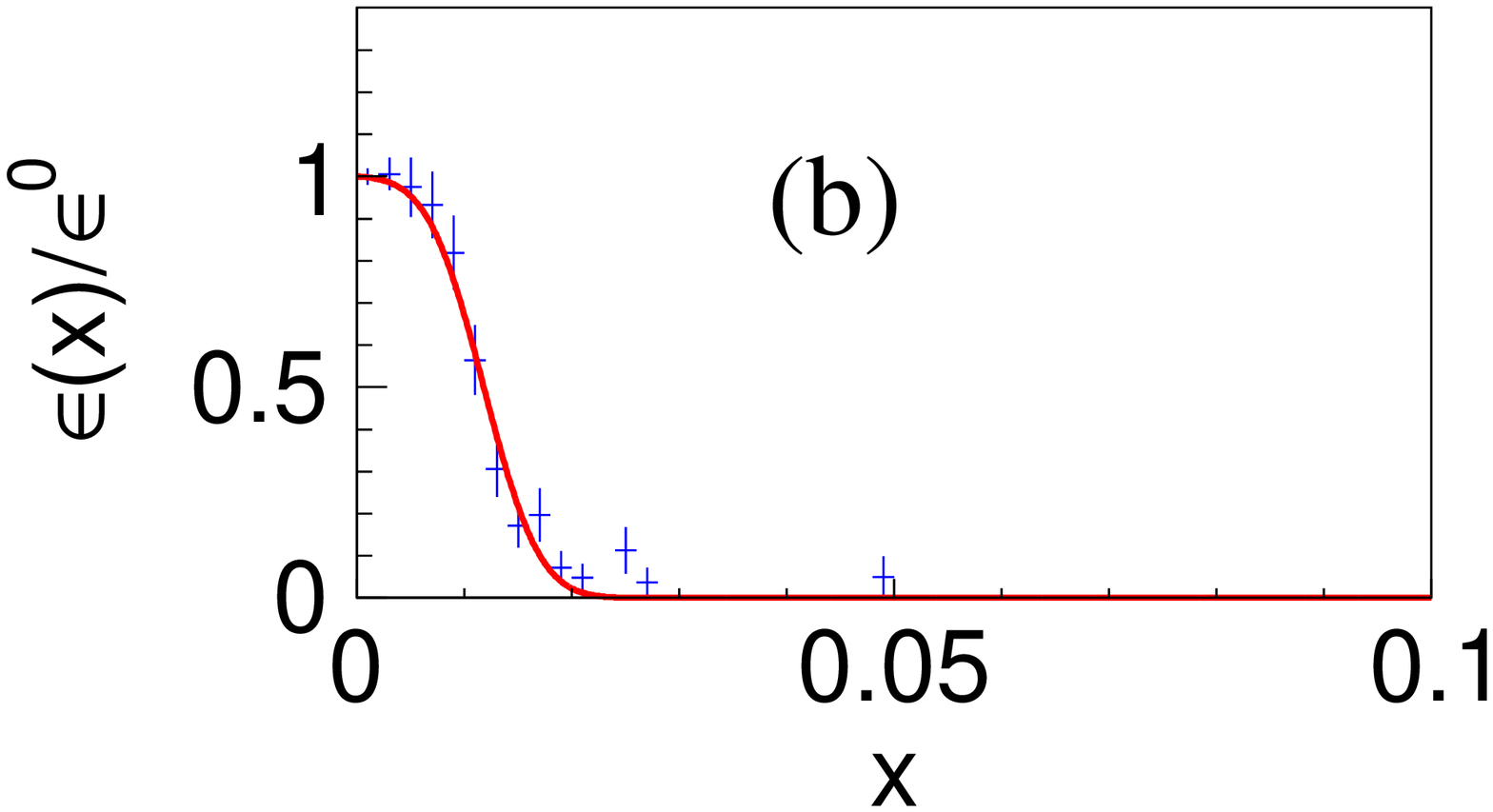}
  \includegraphics[width=0.23\textwidth]{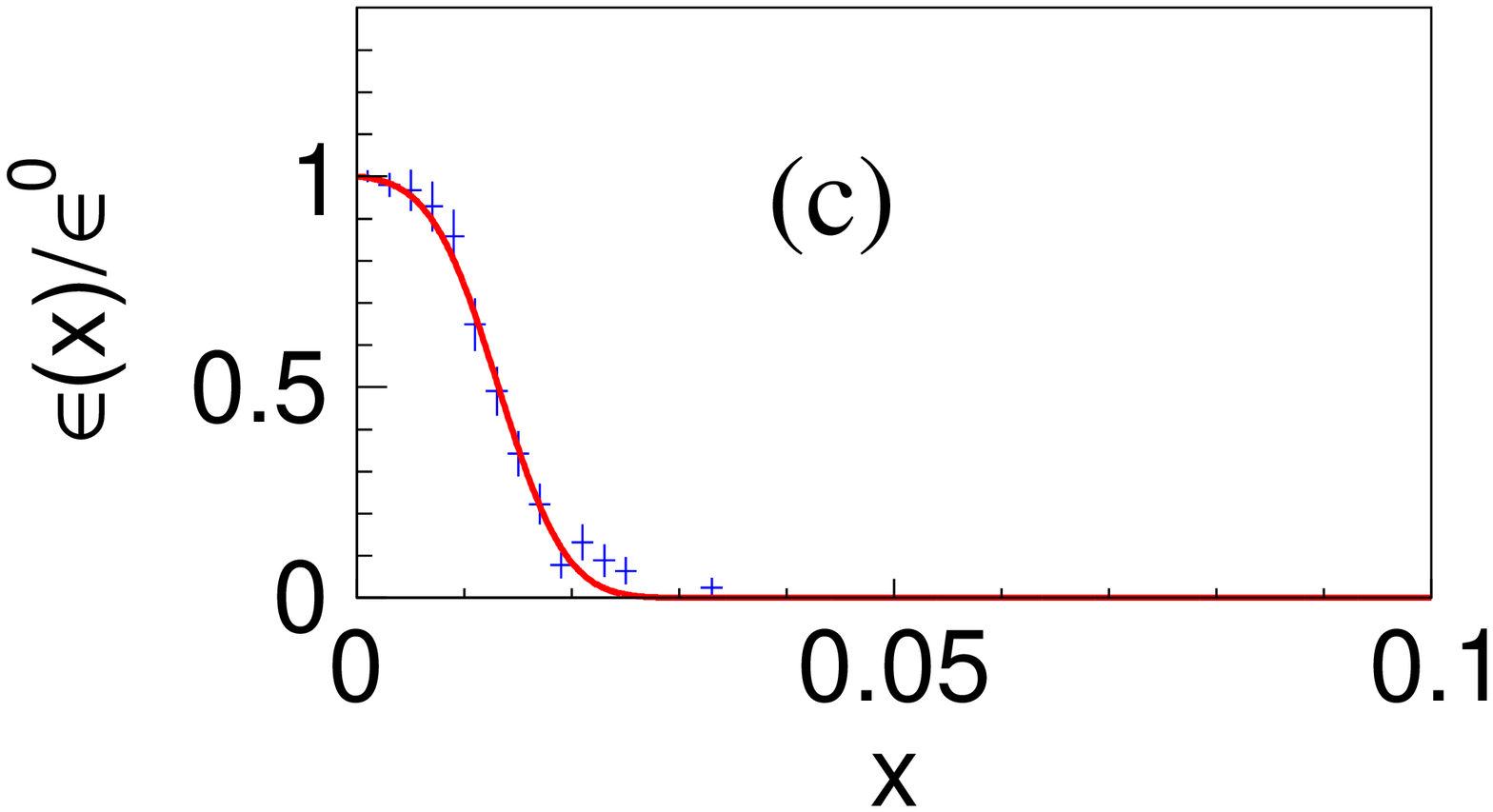}
  \includegraphics[width=0.23\textwidth]{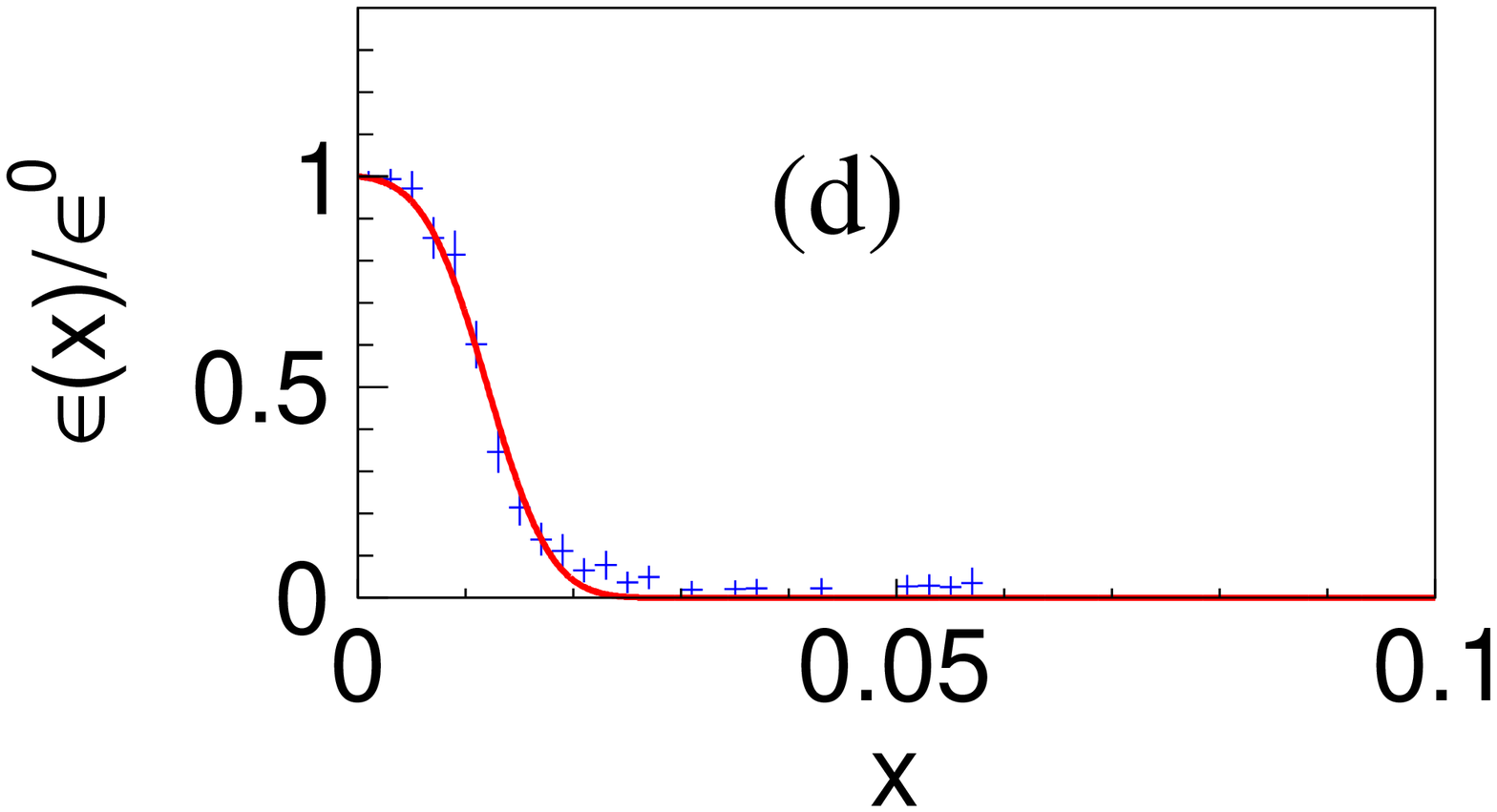}
  \includegraphics[width=0.23\textwidth]{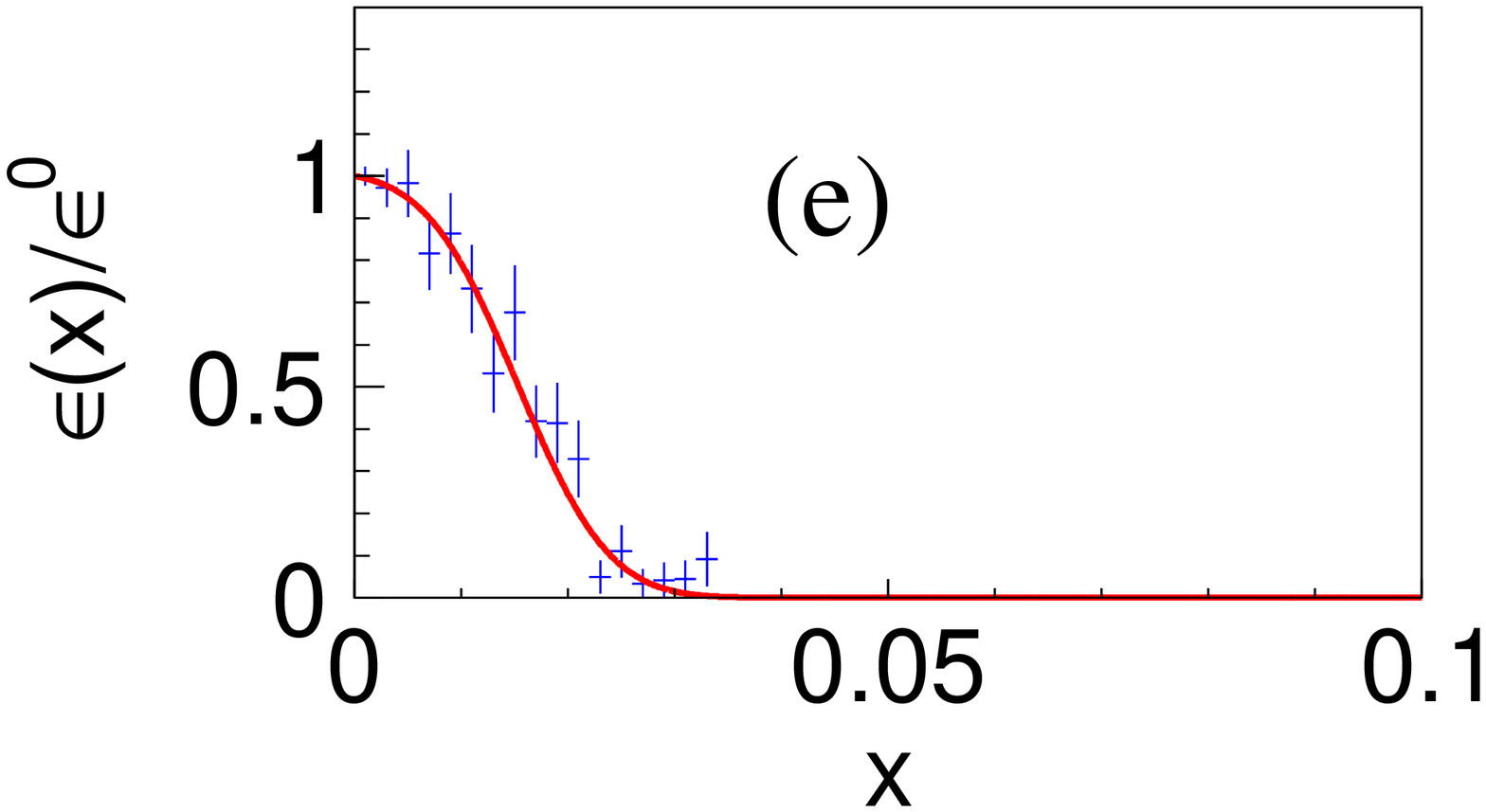}
  \includegraphics[width=0.23\textwidth]{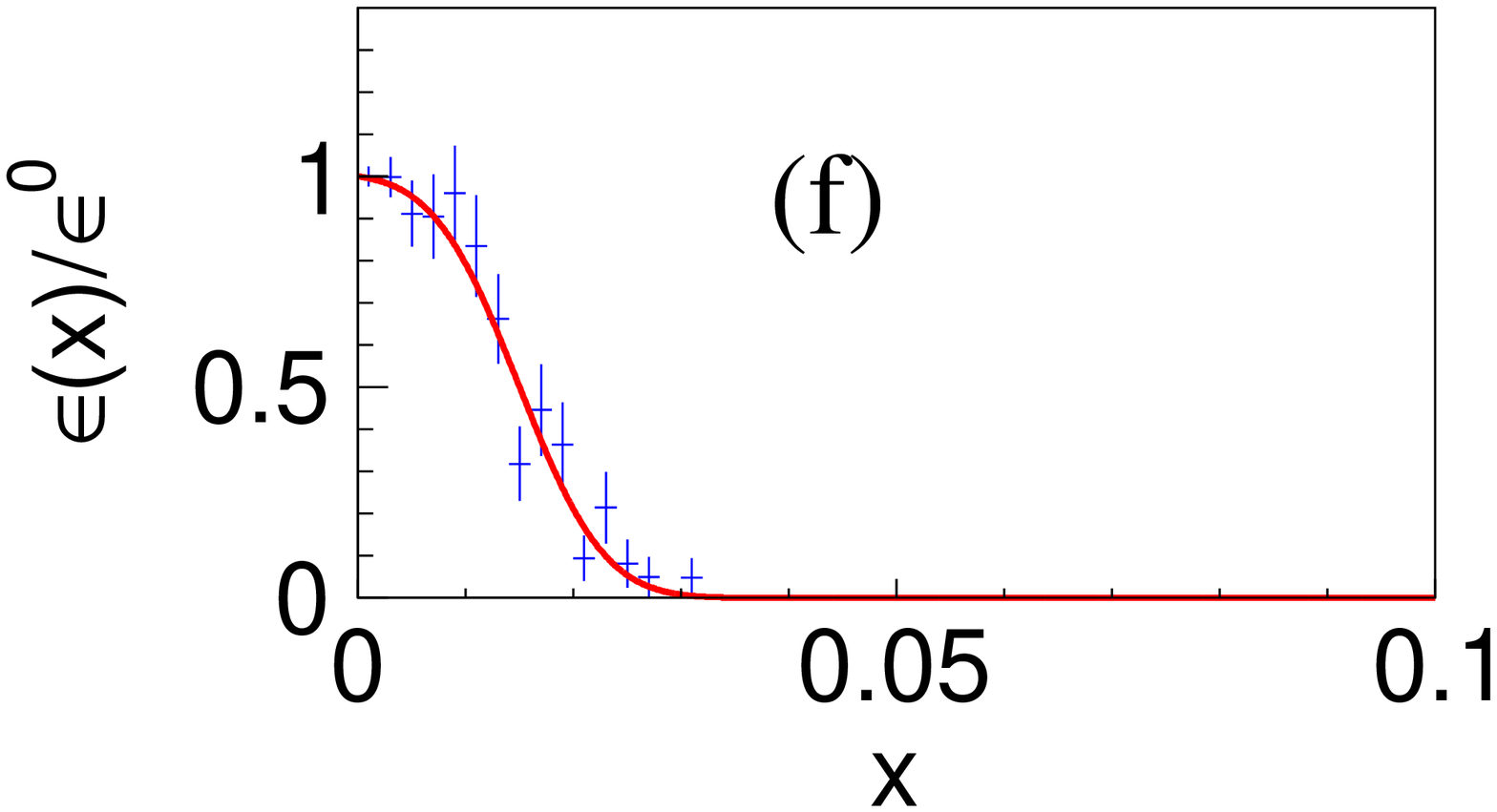}
  \includegraphics[width=0.23\textwidth]{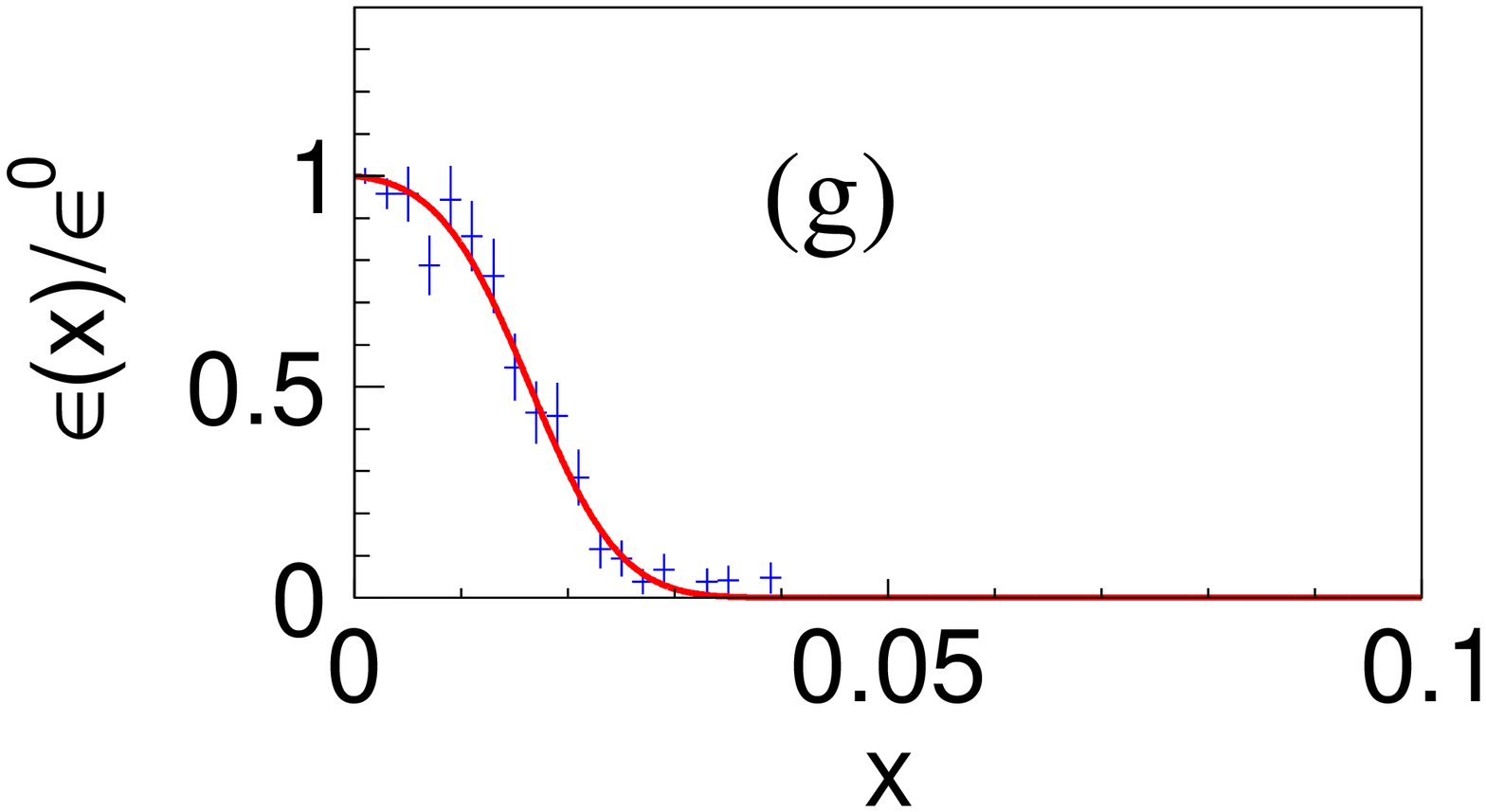}
  \includegraphics[width=0.23\textwidth]{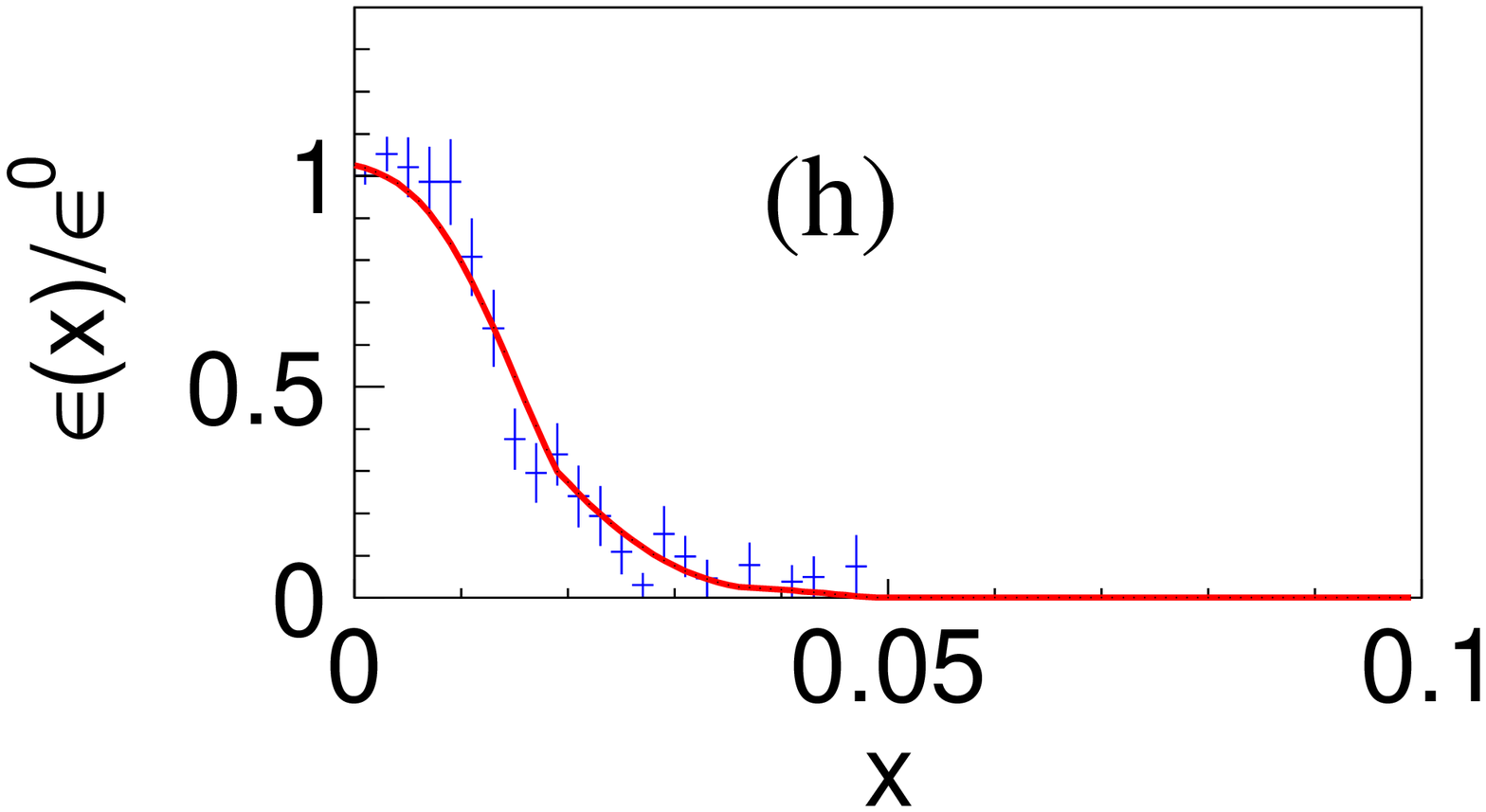}
  \caption{\sf
The distribution of $\epsilon(x)/\epsilon^0$ for (a) $e^+e^-\rightarrow K^+K^-\pi^+\pi^-$, (b) $e^+e^-\rightarrow K^+K^-K^+K^-$, (c) $e^+e^-\rightarrow\pi^+\pi^-\pi^+\pi^-$, (d) $e^+e^-\rightarrow p\bar{p}\pi^+\pi^-$, (e) $e^+e^-\rightarrow K^+K^-\pi^+\pi^-\pi^0$, (f) $e^+e^-\rightarrow K^+K^-K^+K^-\pi^0$, (g) $e^+e^-\rightarrow\pi^+\pi^-\pi^+\pi^-\pi^0$ and (h) $e^+e^-\rightarrow p\bar{p}\pi^+\pi^-\pi^0$ at 4.226 GeV. Points with error bars show the $\epsilon(x)/\epsilon^0$ determined using signal MC. The red lines show the fit to the distributions. To evaluate the goodness of those fits, we check the $\chi^{2}/ndf$ of those plots, which are 0.67, 1.13, 1.13, 1.30, 0.64, 1.00, 0.60, and 1.56, respectively. 
}
  \label{fig:epsilon_x}
\end{figure}
\subsection{Dressed Cross Section}
Inserting the number of observed signal events, the number of peaking and non-peaking backgrounds events, the luminosity of the data sample, the reconstruction efficiency, and the radiative correction factor 
into Eq.~\ref{equation:cs}, we obtain the dressed cross section for $e^+e^-\rightarrow K^+K^-\pi^+\pi^-(\pi^0)$, $K^+K^-K^+K^-(\pi^0)$, $\pi^+\pi^-\pi^+\pi^-(\pi^0)$ and $p\bar{p}\pi^+\pi^-(\pi^0)$ at each energy point. These cross sections are summarized in Table~\ref{table:cs}, where the errors are statistical only.\par 
\begin{table*}[htbp]
\centering
\caption{Summary of the c.m.~energies and luminosities of data sets, together with the dressed cross sections for $e^+e^-\rightarrow K^+K^-\pi^+\pi^-(\pi^0)$, $K^+K^-K^+K^-(\pi^0)$, $\pi^+\pi^-\pi^+\pi^-(\pi^0)$, $p\bar{p}\pi^+\pi^-(\pi^0)$ in the energy region between 3.773 and 4.600 GeV. Errors are statistical only.}
\resizebox{\textwidth}{!}{
\begin{tabular}{|c|c|c|c|c|c|c|c|c|c|} \hline
 \multirow{2}{*}{$E_{\rm cm}$ (GeV)} & \multirow{2}{*}{$L$ (pb$^{-1}$)} & \multicolumn{8}{c|}{$\sigma$ (pb)}\\ 
\cline{3-10}
& & $K^{+}K^{-}\pi^{+}\pi^{-}$ & $K^{+}K^{-}K^{+}K^{-}$ & $\pi^{+}\pi^{-}\pi^{+}\pi^{-}$ & $p\bar{p}\pi^{+}\pi^{-}$ & $K^{+}K^{-}\pi^{+}\pi^{-}\pi^{0}$ & $K^{+}K^{-}K^{+}K^{-}\pi^{0}$ & $\pi^{+}\pi^{-}\pi^{+}\pi^{-}\pi^{0}$& $p\bar{p}\pi^{+}\pi^{-}\pi^{0}$\\
\hline                                           
3.773&    2931.8&    $122.2\pm0.4$&  $21.4\pm0.2$& $172.0\pm0.4$& $38.9\pm0.2$& $ 191.2\pm 0.6$& $ 7.2\pm 0.2$&  $77.0\pm0.4$& $ 59.6\pm 0.3$ \\
3.808&      50.5&    $120.2\pm3.1$&  $19.9\pm1.6$& $173.0\pm3.3$& $37.0\pm1.4$& $ 185.5\pm 4.5$& $ 6.5\pm 1.4$&  $80.0\pm3.1$& $ 57.2\pm 2.5$ \\
3.867&     108.9&    $104.5\pm2.0$&  $18.0\pm1.0$& $155.4\pm2.1$& $34.9\pm0.9$& $ 166.5\pm 2.9$& $ 7.8\pm 1.0$&  $70.4\pm2.0$& $ 55.6\pm 1.6$ \\
3.871&     110.3&    $104.3\pm2.0$&  $16.3\pm0.9$& $150.3\pm2.1$& $34.3\pm0.9$& $ 165.4\pm 2.9$& $ 7.4\pm 1.0$&  $71.1\pm2.0$& $ 55.3\pm 1.6$ \\
3.896&      52.6&    $104.8\pm2.8$&  $19.0\pm1.5$& $147.3\pm3.0$& $33.2\pm1.3$& $ 165.4\pm 4.2$& $ 6.5\pm 1.3$&  $64.1\pm2.7$& $ 52.3\pm 2.3$ \\
4.008&     482.0&    $ 85.2\pm0.8$&  $15.4\pm0.5$& $119.8\pm0.9$& $26.1\pm0.4$& $ 144.0\pm 1.3$& $ 6.6\pm 0.4$&  $62.4\pm0.9$& $ 48.7\pm 0.7$ \\
4.085&      52.9&    $ 72.1\pm2.3$&  $13.2\pm1.2$& $104.6\pm2.6$& $21.9\pm1.0$& $ 132.8\pm 3.7$& $ 3.8\pm 1.3$&  $56.3\pm2.5$& $ 42.9\pm 2.1$ \\
4.129&     393.4&    $ 67.6\pm0.8$&  $12.5\pm0.5$& $ 99.5\pm0.9$& $21.8\pm0.4$& $ 119.4\pm 1.3$& $ 5.9\pm 0.5$&  $48.4\pm0.8$& $ 41.2\pm 0.7$ \\
4.158&     406.9&    $ 66.5\pm0.7$&  $13.0\pm0.5$& $ 94.8\pm0.9$& $20.8\pm0.4$& $ 115.1\pm 1.2$& $ 6.6\pm 0.5$&  $45.7\pm0.8$& $ 40.2\pm 0.7$ \\
4.178&    3194.5&    $ 66.1\pm0.3$&  $12.6\pm0.1$& $ 91.2\pm0.3$& $20.3\pm0.1$& $ 116.8\pm 0.5$& $ 5.9\pm 0.2$&  $44.8\pm0.3$& $ 40.3\pm 0.3$ \\
4.189&      43.3&    $ 65.0\pm2.4$&  $12.6\pm1.3$& $ 93.3\pm2.7$& $21.6\pm1.1$& $ 111.0\pm 3.7$& $ 5.0\pm 1.5$&  $44.5\pm2.5$& $ 37.5\pm 2.1$ \\
4.189&     524.6&    $ 66.0\pm0.7$&  $11.5\pm0.3$& $ 92.2\pm0.7$& $19.9\pm0.3$& $ 112.1\pm 1.1$& $ 5.7\pm 0.4$&  $43.8\pm0.7$& $ 39.6\pm 0.6$ \\
4.199&     526.0&    $ 62.6\pm0.7$&  $11.8\pm0.3$& $ 87.5\pm0.7$& $19.4\pm0.3$& $ 113.8\pm 1.1$& $ 5.7\pm 0.4$&  $43.1\pm0.7$& $ 39.1\pm 0.6$ \\
4.208&      55.0&    $ 60.9\pm2.1$&  $11.1\pm1.0$& $ 93.4\pm2.3$& $19.8\pm1.0$& $ 115.9\pm 3.4$& $ 4.9\pm 1.2$&  $44.8\pm2.2$& $ 40.9\pm 1.9$ \\
4.209&     518.0&    $ 63.4\pm0.7$&  $12.2\pm0.3$& $ 85.3\pm0.7$& $19.6\pm0.3$& $ 108.9\pm 1.1$& $ 5.3\pm 0.4$&  $41.8\pm0.7$& $ 38.7\pm 0.6$ \\
4.217&      54.6&    $ 62.6\pm2.1$&  $11.4\pm1.0$& $ 85.0\pm2.2$& $20.8\pm1.0$& $ 112.6\pm 3.3$& $ 5.6\pm 1.2$&  $44.4\pm2.2$& $ 38.4\pm 1.9$ \\
4.219&     514.6&    $ 60.8\pm0.7$&  $12.1\pm0.3$& $ 88.8\pm0.7$& $18.8\pm0.3$& $ 109.3\pm 1.1$& $ 5.1\pm 0.4$&  $41.1\pm0.7$& $ 36.7\pm 0.6$ \\
4.226&      44.5&    $ 65.0\pm2.4$&  $13.5\pm1.3$& $ 85.6\pm2.4$& $18.4\pm1.0$& $ 114.3\pm 3.7$& $ 5.6\pm 1.3$&  $43.1\pm2.4$& $ 40.4\pm 2.1$ \\
4.226&    1056.4&    $ 62.4\pm0.5$&  $11.9\pm0.2$& $ 88.1\pm0.5$& $18.2\pm0.2$& $ 109.9\pm 0.7$& $ 5.0\pm 0.3$&  $42.8\pm0.5$& $ 38.2\pm 0.4$ \\
4.236&     530.3&    $ 59.6\pm0.6$&  $11.8\pm0.3$& $ 86.1\pm0.7$& $19.4\pm0.3$& $ 109.1\pm 1.1$& $ 5.5\pm 0.4$&  $41.4\pm0.7$& $ 38.2\pm 0.6$ \\
4.242&      55.9&    $ 59.8\pm2.0$&  $10.7\pm1.0$& $ 84.2\pm2.2$& $17.4\pm0.9$& $ 104.9\pm 3.2$& $ 4.7\pm 1.1$&  $41.0\pm2.1$& $ 37.3\pm 1.8$ \\
4.244&     538.1&    $ 59.1\pm0.6$&  $11.8\pm0.3$& $ 82.6\pm0.7$& $18.9\pm0.3$& $ 106.0\pm 1.0$& $ 5.2\pm 0.4$&  $41.5\pm0.7$& $ 36.7\pm 0.6$ \\
4.258&     828.4&    $ 58.6\pm0.5$&  $11.3\pm0.2$& $ 83.1\pm0.5$& $18.0\pm0.2$& $ 105.9\pm 0.8$& $ 4.7\pm 0.3$&  $40.8\pm0.5$& $ 38.3\pm 0.5$ \\
4.267&     531.1&    $ 55.1\pm0.6$&  $11.1\pm0.3$& $ 81.6\pm0.7$& $18.0\pm0.3$& $ 104.1\pm 1.0$& $ 4.6\pm 0.4$&  $39.7\pm0.6$& $ 36.7\pm 0.6$ \\
4.278&     175.7&    $ 55.4\pm1.1$&  $11.2\pm0.6$& $ 80.2\pm1.3$& $16.8\pm0.5$& $ 102.0\pm 1.8$& $ 3.9\pm 0.6$&  $38.3\pm1.1$& $ 35.6\pm 1.0$ \\
4.288&     491.5&    $ 55.1\pm0.6$&  $10.5\pm0.3$& $ 77.1\pm0.7$& $17.3\pm0.3$& $  99.1\pm 1.0$& $ 6.0\pm 0.4$&  $37.9\pm0.7$& $ 34.7\pm 0.6$ \\
4.308&      45.1&    $ 53.3\pm2.1$&  $10.9\pm1.2$& $ 77.3\pm2.3$& $17.1\pm1.0$& $  95.1\pm 3.3$& $ 3.5\pm 1.2$&  $38.0\pm2.3$& $ 33.8\pm 1.9$ \\
4.312&     492.1&    $ 53.0\pm0.6$&  $10.2\pm0.3$& $ 74.4\pm0.7$& $16.8\pm0.3$& $  95.3\pm 1.0$& $ 5.4\pm 0.4$&  $37.7\pm0.6$& $ 34.9\pm 0.6$ \\
4.338&     501.1&    $ 50.6\pm0.6$&  $ 9.8\pm0.3$& $ 72.7\pm0.6$& $15.9\pm0.3$& $  94.4\pm 1.0$& $ 5.2\pm 0.4$&  $38.3\pm0.6$& $ 34.2\pm 0.6$ \\
4.358&     543.9&    $ 50.8\pm0.6$&  $10.0\pm0.3$& $ 71.2\pm0.6$& $15.0\pm0.3$& $  92.8\pm 0.9$& $ 4.6\pm 0.4$&  $36.7\pm0.6$& $ 32.9\pm 0.5$ \\
4.378&     522.8&    $ 46.7\pm0.6$&  $ 9.8\pm0.3$& $ 70.1\pm0.6$& $14.8\pm0.3$& $  90.2\pm 0.9$& $ 5.0\pm 0.4$&  $36.4\pm0.6$& $ 33.6\pm 0.6$ \\
4.387&      55.6&    $ 46.7\pm1.8$&  $10.2\pm0.9$& $ 67.8\pm2.0$& $14.5\pm0.8$& $  93.5\pm 3.0$& $ 4.4\pm 1.1$&  $37.0\pm2.0$& $ 33.4\pm 1.7$ \\
4.397&     505.0&    $ 46.2\pm0.6$&  $ 9.2\pm0.3$& $ 67.1\pm0.6$& $14.4\pm0.3$& $  90.0\pm 1.0$& $ 5.2\pm 0.4$&  $34.8\pm0.6$& $ 30.8\pm 0.6$ \\
4.416&      46.8&    $ 48.0\pm2.0$&  $ 9.8\pm1.0$& $ 66.5\pm2.1$& $13.1\pm0.8$& $  87.4\pm 3.1$& $ 4.7\pm 1.2$&  $34.7\pm2.1$& $ 32.6\pm 1.8$ \\
4.416&    1043.9&    $ 46.6\pm0.4$&  $ 8.8\pm0.2$& $ 63.4\pm0.4$& $14.3\pm0.2$& $  87.8\pm 0.7$& $ 4.4\pm 0.2$&  $33.0\pm0.4$& $ 32.3\pm 0.4$ \\
4.437&     568.1&    $ 46.0\pm0.5$&  $ 9.3\pm0.2$& $ 64.1\pm0.6$& $14.2\pm0.2$& $  86.0\pm 0.9$& $ 5.0\pm 0.3$&  $32.0\pm0.6$& $ 31.1\pm 0.5$ \\
4.467&     111.1&    $ 41.5\pm1.1$&  $ 8.5\pm0.6$& $ 60.3\pm1.4$& $13.4\pm0.5$& $  83.5\pm 2.0$& $ 4.1\pm 0.7$&  $31.1\pm1.2$& $ 29.7\pm 1.1$ \\
4.527&     112.1&    $ 41.7\pm1.1$&  $ 7.2\pm0.6$& $ 55.7\pm1.3$& $12.9\pm0.5$& $  77.6\pm 1.9$& $ 3.4\pm 0.7$&  $29.4\pm1.2$& $ 26.3\pm 1.1$ \\
4.575&      48.9&    $ 35.8\pm1.7$&  $ 6.5\pm0.8$& $ 50.2\pm1.8$& $11.3\pm0.8$& $  73.5\pm 2.8$& $ 3.2\pm 1.1$&  $25.4\pm1.8$& $ 28.7\pm 1.7$ \\
4.600&     586.9&    $ 36.2\pm0.5$&  $ 7.1\pm0.2$& $ 49.4\pm0.5$& $10.9\pm0.2$& $  72.4\pm 0.8$& $ 3.6\pm 0.3$&  $26.5\pm0.5$& $ 25.8\pm 0.5$ \\
\hline                                                                                       
\end{tabular}                                                                                
}
\label{table:cs}
\end{table*}
\subsection{Analysis of Dressed Cross Section}
A least $\chi^{2}$ fit is applied to those dressed cross sections. The function to be minimized is
\begin{equation}
\chi^{2}=\sum_{i=1}^{N}\left(\frac{\sigma(E_{\rm{cm}})-\sigma^{\rm expected}(E_{\rm{cm}})}{\Delta_{\sigma(E_{\rm{cm}})}}\right)^{2},
\end{equation}
where $\sigma^{\rm expected}(E_{\rm{cm}})$ is the theoretically expected dressed cross section, $\sigma(E_{\rm{cm}})$ is the measured value of the dressed cross section with combined statistical and uncorrelated systematic error $\Delta_{\sigma(E_{\rm{cm}})}$ at the $i^{\rm th}$ energy point, and $N$ is the number of data samples collected at different energy points.\par
First, we consider the expected dressed cross section of the continuum process:
\begin{equation}
\sigma^{\rm expected}_{\rm cont}=|A_{\rm cont}|^{2},
\end{equation}
where $A_{\rm cont}$ is the amplitude of the continuum process,
\begin{equation}
A_{\rm cont}=\sqrt{\frac{f_{\rm cont}}{(E_{\rm cm}/4.226\,{\rm GeV})^{n}}},
\end{equation}
and where $E_{\rm cm}$ is the c.m.~energy, $f_{\rm cont}$ and $n$ are floating parameters in the fit. Results from these fits, which only consider the contribution from the continuum process, are shown in Fig.~\ref{fig:cs_fit_results} and listed in Table~\ref{table:fit_cont}.
\begin{table}[htbp]
\centering
\caption{Summary of fit results considering only the contribution from the continuum process. The first errors are statistical and the second systematic.}
\resizebox{0.50\textwidth}{!}{
\begin{tabular}{|c|c|c|c|} \hline
Final state                 & $\chi^{2}/ndf$ & $f_{\rm cont}$ (GeV$^{n}$/pb) &  $n$ \\
\hline
$K^+K^-\pi^+\pi^-$         & 1.24 & $ 60.65\pm0.19\pm2.97$ & $6.24\pm0.07\pm0.01$\\
$K^+K^-K^+K^-$             & 0.78 &  $11.59\pm0.08\pm0.57$ & $5.36\pm0.14\pm0.01$\\
$\pi^+\pi^-\pi^+\pi^-$     & 1.11 &  $85.94\pm0.25\pm4.38$ & $6.36\pm0.06\pm0.01$\\
$p\bar{p}\pi^+\pi^-$       & 0.89 &  $18.88\pm0.08\pm0.83$ & $6.43\pm0.09\pm0.01$\\
$K^+K^-\pi^+\pi^-\pi^0$    & 1.11 & $108.72\pm0.25\pm5.11$ & $4.97\pm0.05\pm0.01$\\
$K^+K^-K^+K^-\pi^0$        & 1.07 &   $5.29\pm0.07\pm0.33$ & $3.08\pm0.26\pm0.01$\\
$\pi^+\pi^-\pi^+\pi^-\pi^0$& 1.06 &  $42.44\pm0.21\pm2.37$ & $5.54\pm0.11\pm0.01$\\
$p\bar{p}\pi^+\pi^-\pi^0$  & 1.23 &  $37.91\pm0.12\pm1.74$ & $4.11\pm0.07\pm0.01$\\
\hline
\end{tabular}
}
\label{table:fit_cont}
\end{table}
\begin{figure}[htbp]
  \centering
  \includegraphics[width=0.23\textwidth]{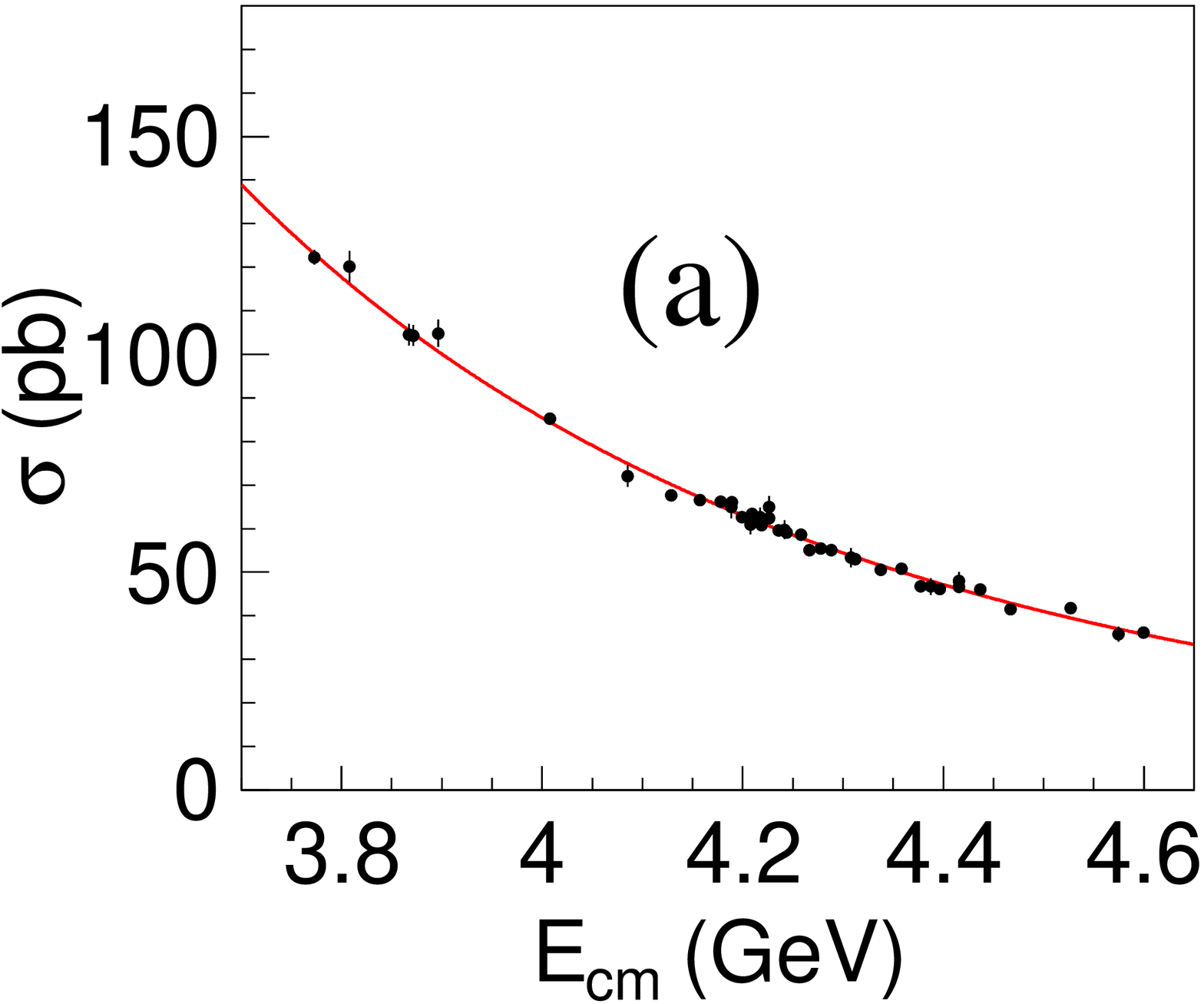}     
  \includegraphics[width=0.23\textwidth]{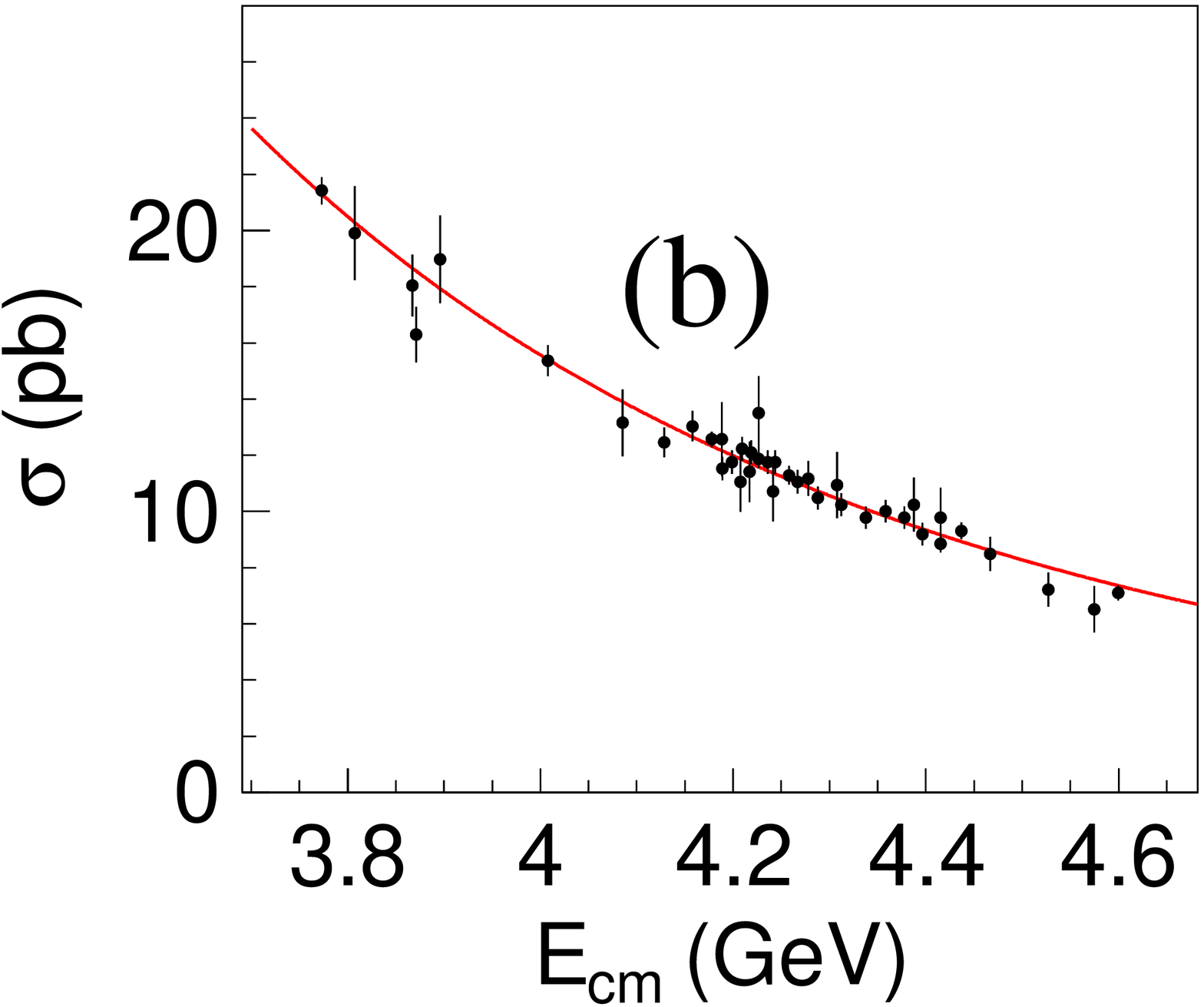}        
  \includegraphics[width=0.23\textwidth]{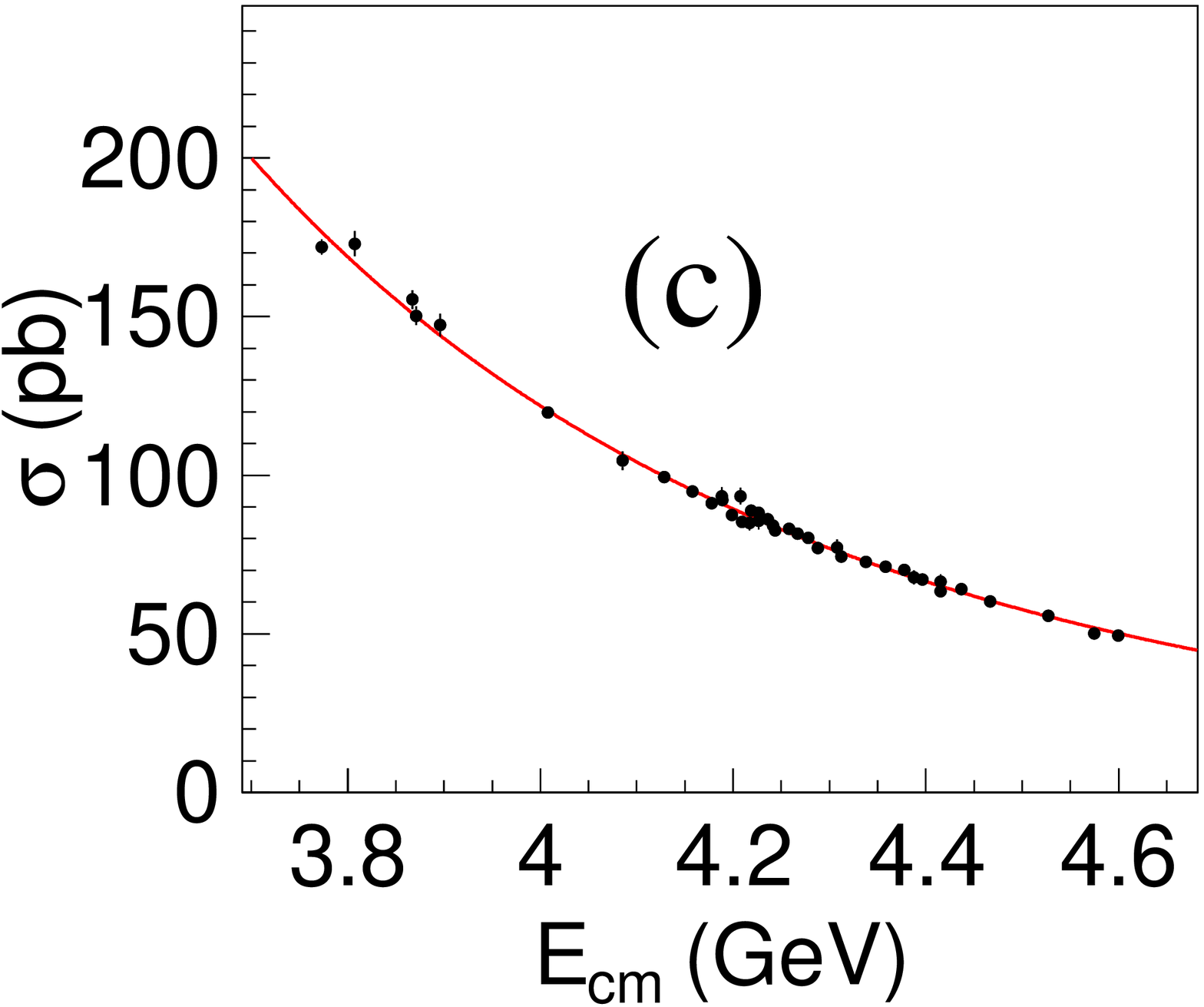}       
  \includegraphics[width=0.23\textwidth]{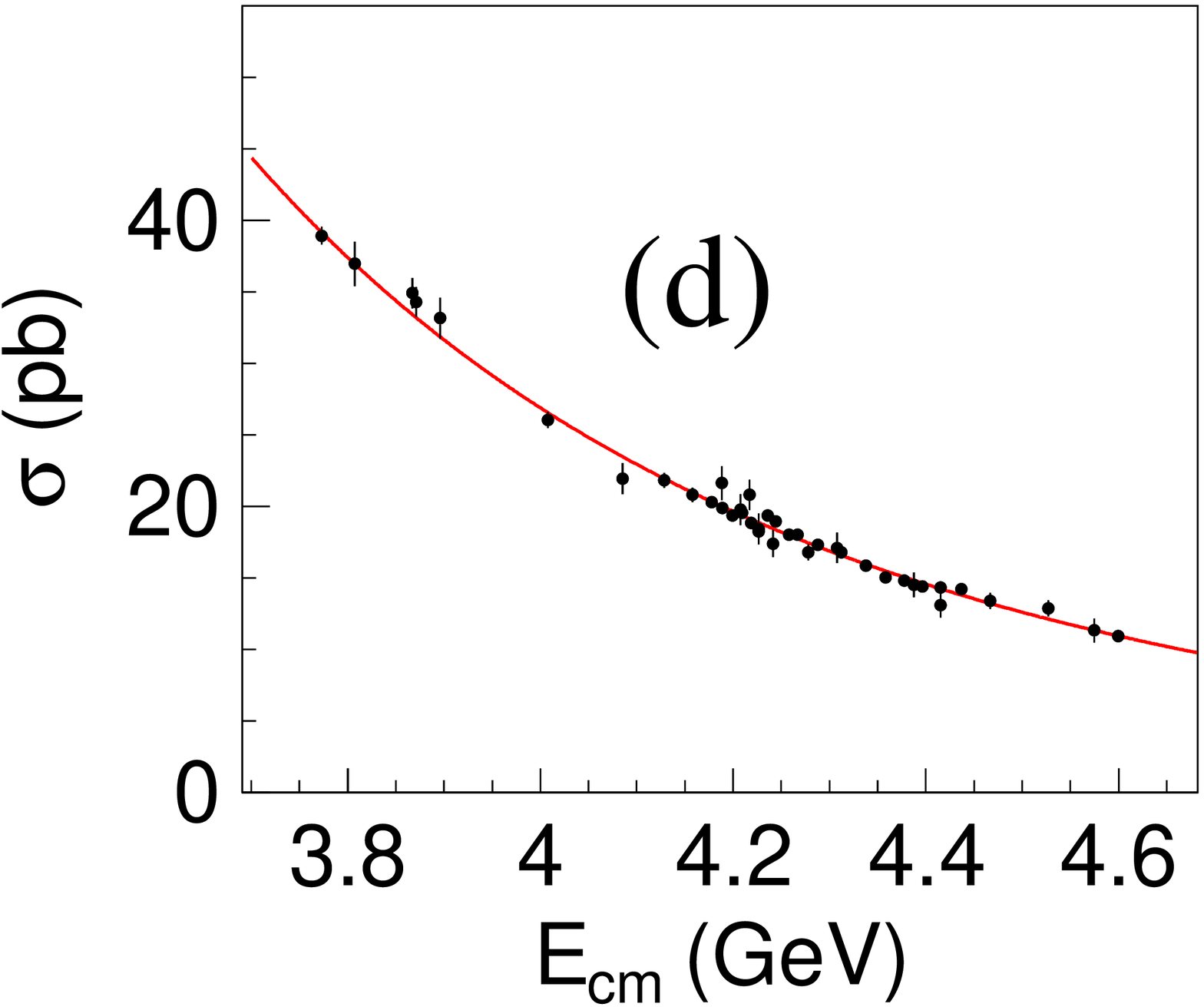}     
  \includegraphics[width=0.23\textwidth]{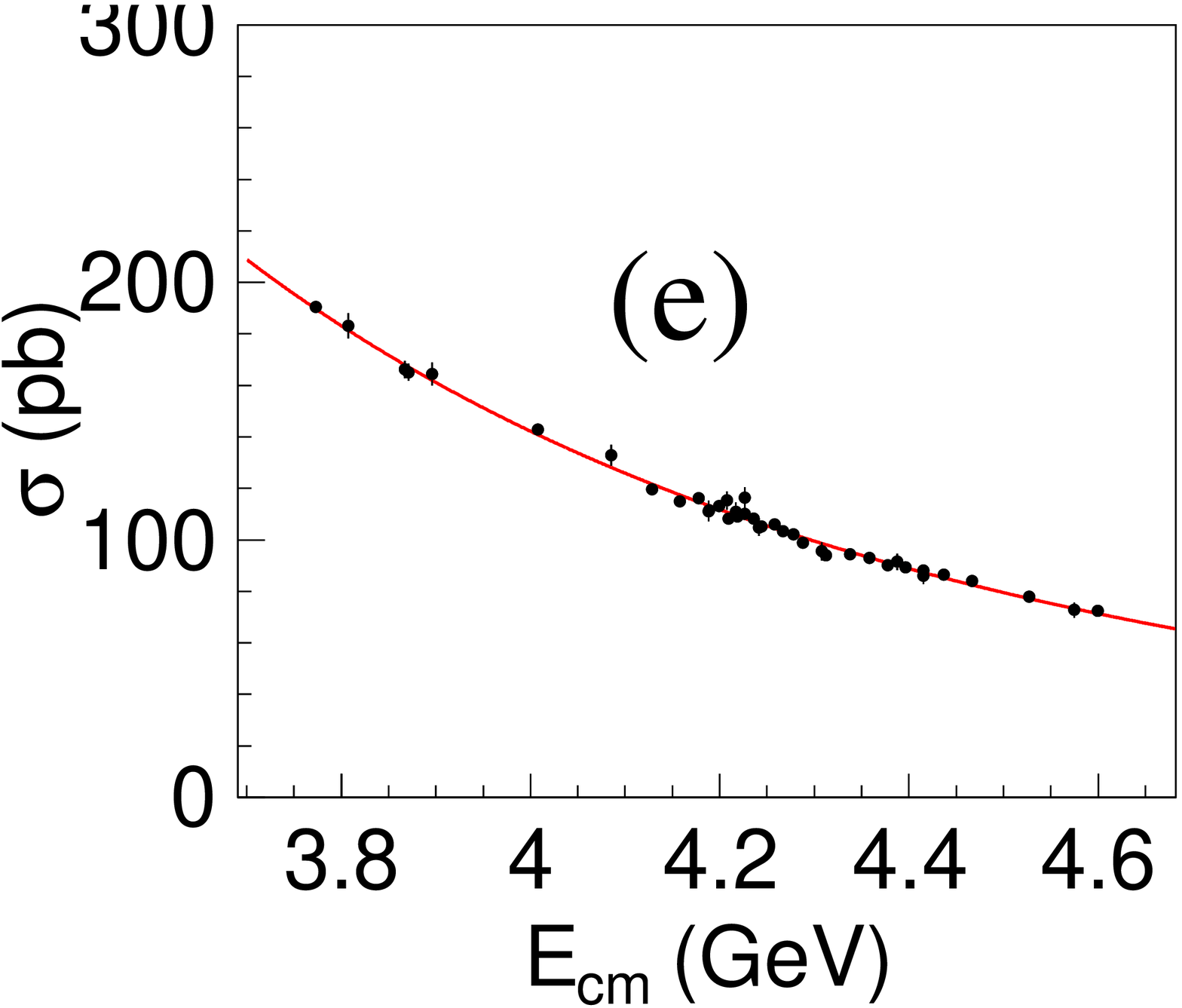}  
  \includegraphics[width=0.23\textwidth]{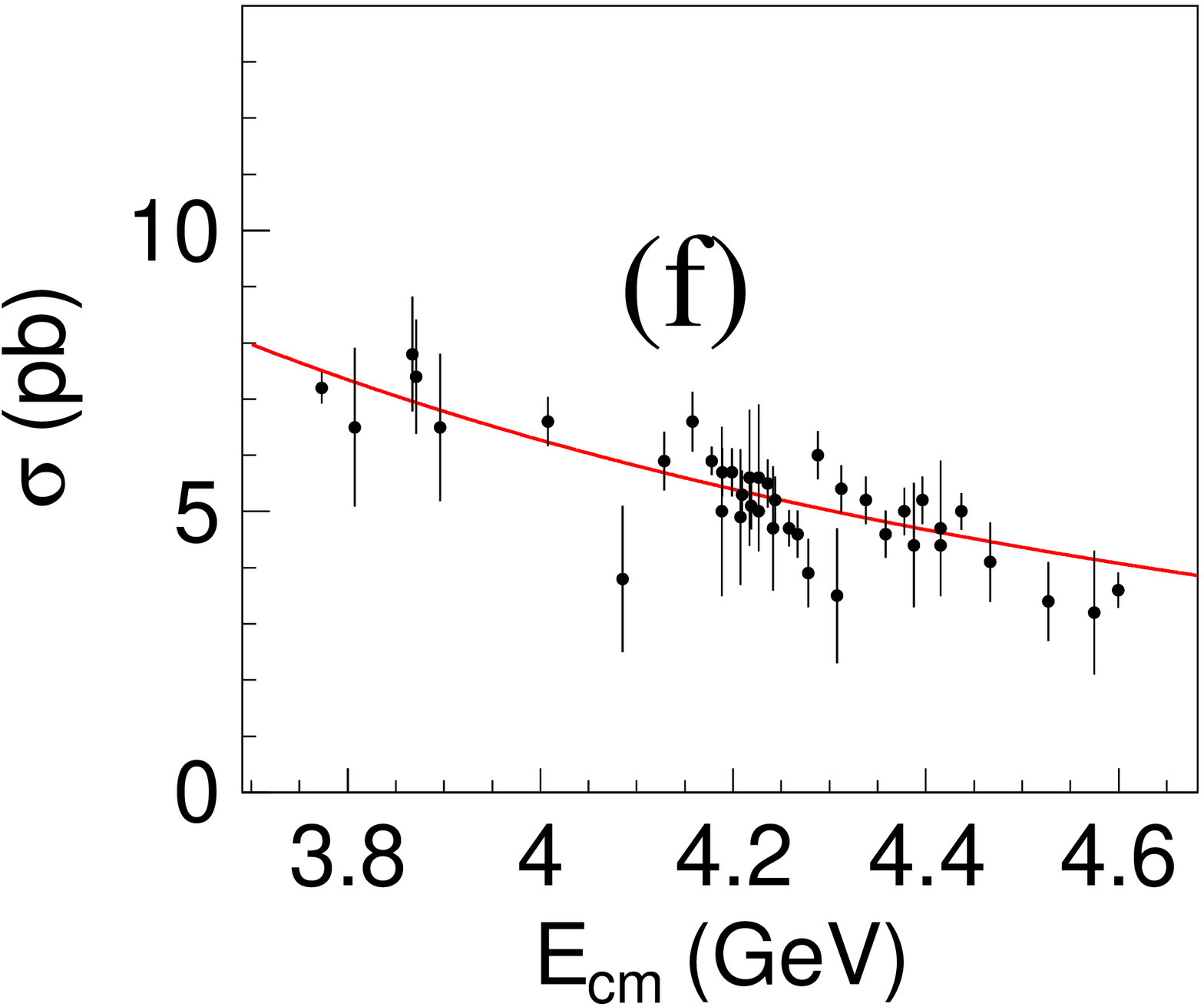}     
  \includegraphics[width=0.23\textwidth]{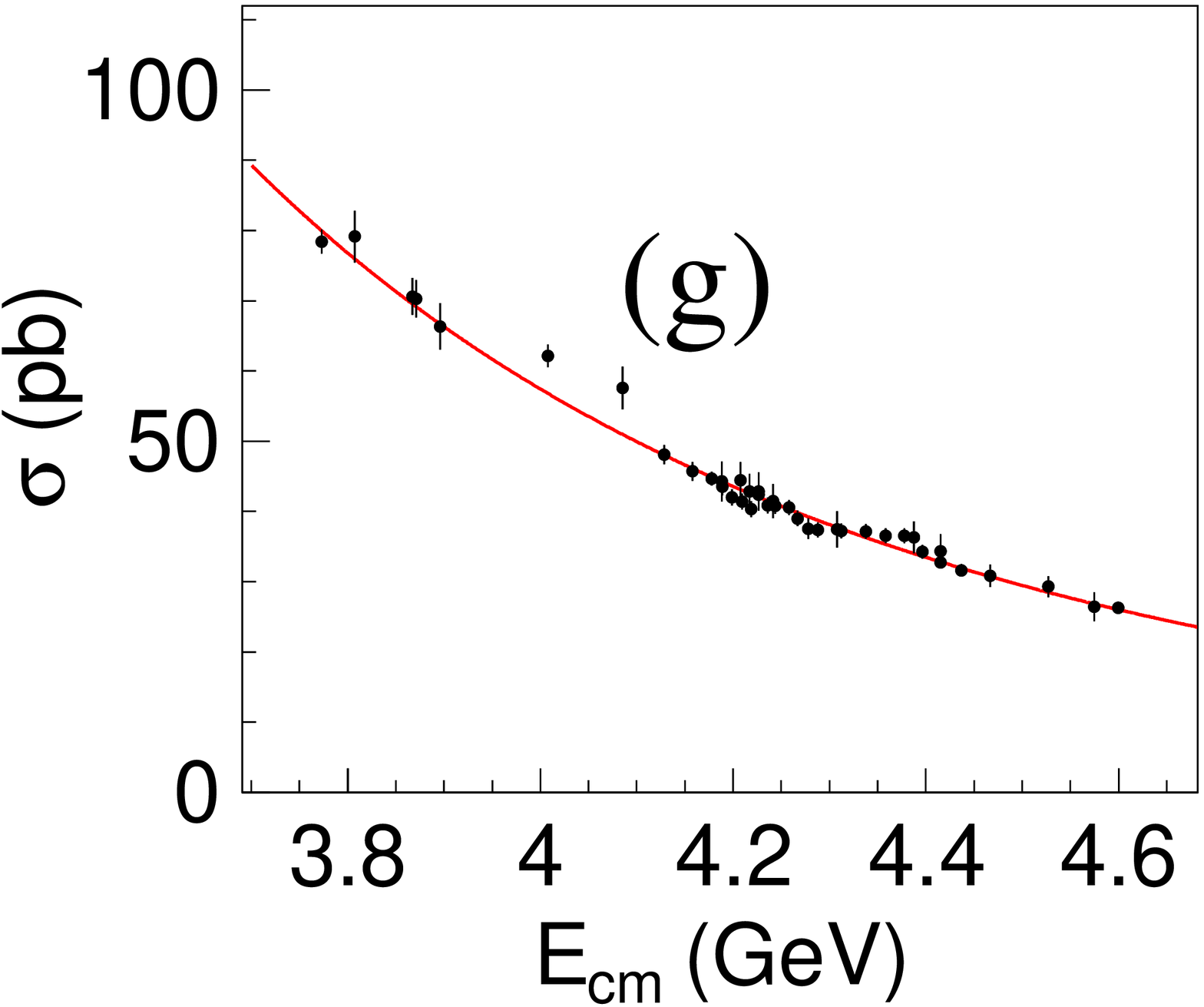}    
  \includegraphics[width=0.23\textwidth]{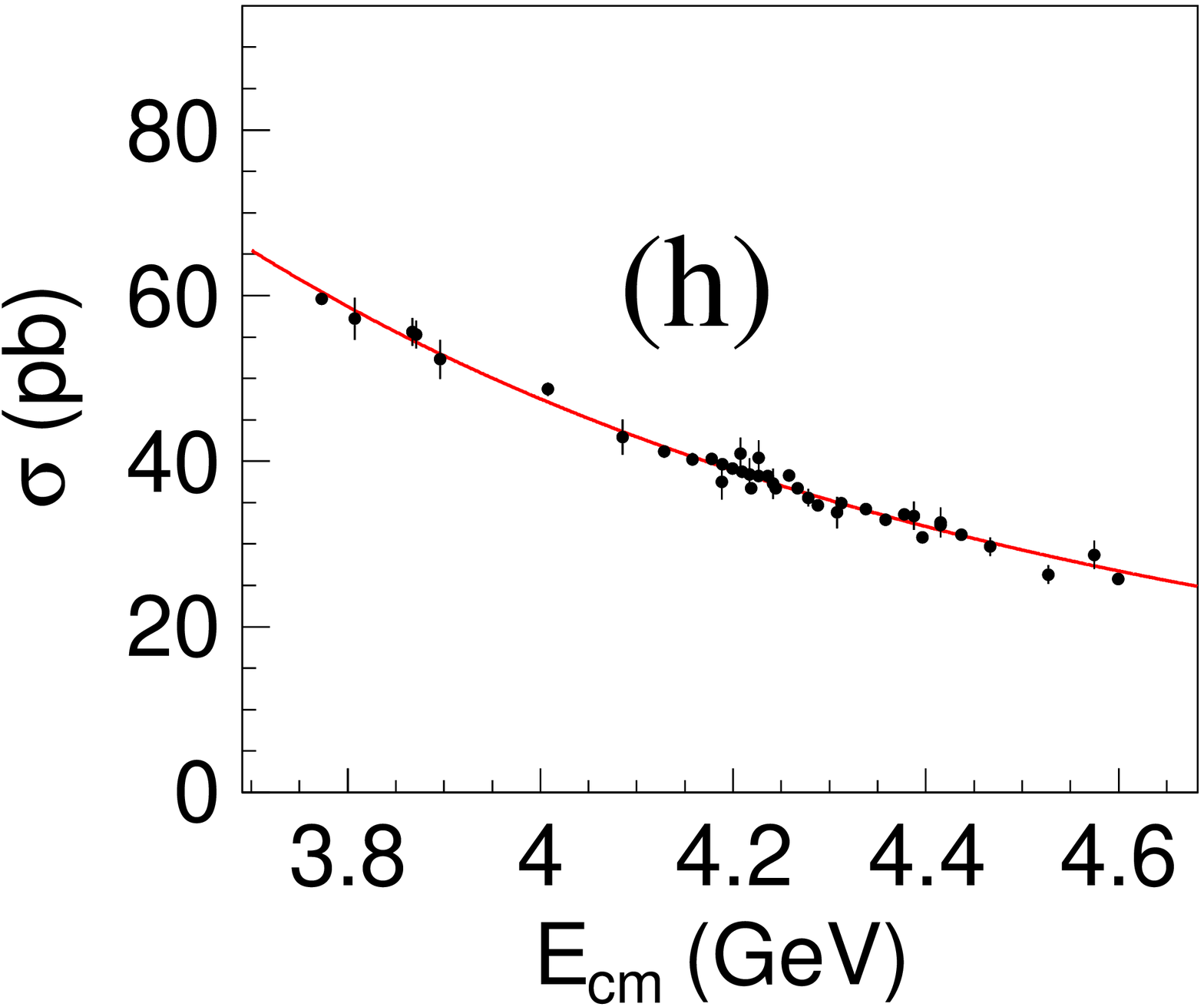}  
  \caption{\sf
Fits to dressed cross sections for (a) $e^+e^-\rightarrow K^+K^-\pi^+\pi^-$, (b) $e^+e^-\rightarrow K^+K^-K^+K^-$, (c) $e^+e^-\rightarrow\pi^+\pi^-\pi^+\pi^-$, (d) $e^+e^-\rightarrow p\bar{p}\pi^+\pi^-$, (e) $e^+e^-\rightarrow K^+K^-\pi^+\pi^-\pi^0$, (f) $e^+e^-\rightarrow K^+K^-K^+K^-\pi^0$, (g) $e^+e^-\rightarrow\pi^+\pi^-\pi^+\pi^-\pi^0$ and (h) $e^+e^-\rightarrow p\bar{p}\pi^+\pi^-\pi^0$ only considering contribution from continuum process. Points with error bars show the measured dressed cross sections. The red lines show the fit results.
}
  \label{fig:cs_fit_results}
\end{figure}
\subsubsection{Branching fraction of $\psi(4040)\rightarrow\pi^+\pi^-\pi^+\pi^-\pi^0$}
For the dressed cross section of $e^+e^-\rightarrow\pi^+\pi^-\pi^+\pi^-\pi^0$, we also constructed a fit amplitude including a contribution from $\psi(4040)$ decay:
\begin{equation}
\sigma^{\rm expected}_{\rm{cont}+\psi(4040)}=|A_{\rm cont}+A_{\psi(4040)}e^{i\phi}|^{2},
\end{equation}
where $\phi$ is the relative phase, and $A_{\psi(4040)}$ is a Breit-Wigner for the production of the $\psi(4040)$:
\begin{equation} 
\footnotesize
A_{\psi(4040)}=\frac{\sqrt{12\pi\Gamma^{ee}_{\psi(4040)}\Gamma^{\rm total}_{\psi(4040)}{\rm BF}(\psi(4040)\rightarrow\pi^+\pi^-\pi^+\pi^-\pi^0)}}{(E_{\rm cm}^{2}-M^2_{\psi(4040)})+i\Gamma^{\rm total}_{\psi(4040)}M_{\psi(4040)}},
\end{equation} 
where $M_{\psi(4040)}$, $\Gamma^{\rm total}_{\psi(4040)}$ and $\Gamma^{ee}_{\psi(4040)}$ are the mass, the total width, and the leptonic width of the $\psi(4040)$. In this fit, the values of $\Gamma^{\rm total}_{\psi(4040)}$ and $\Gamma^{ee}_{\psi(4040)}$ are fixed according to the PDG listings~\cite{PDG}.
The fit has two solutions with equally good fit quality, while
the phases and the branching fractions are different.
The fit is shown in Fig.~\ref{fig:cs_fit_results_1stru}, 
and the fit parameters are listed in Table~\ref{table:fit_cont_4040}.\par
\begin{table}[htbp]
\centering
\caption{Summary of fit results for the $\pi^+\pi^-\pi^+\pi^-\pi^0$ final state, including the continuum process, $\psi(4040)$ decay, and interference between them. The first errors are statistical and the second systematic.}
\resizebox{0.45\textwidth}{!}{
\begin{tabular}{|c|c|} \hline
Parameter  & Solution 1  \\
\hline                                           
$\chi^{2}/ndf$                            &      0.67         \\ 
$f_{\rm cont}$                            & $(9.13\pm1.67\pm0.72)\times10^{4}$ GeV$^{n}$/pb     \\
$n$                                                       & $5.33\pm0.13\pm0.02$                \\
$\rm{BF}(\psi(4040)\rightarrow\pi^+\pi^-\pi^+\pi^-\pi^0)$ & $(3.51\pm1.89\pm1.24)\times10^{-5}$ \\
$\phi$                                                    & $(109.73\pm14.04\pm3.81)^{\circ}$   \\
\hline
Parameter  & Solution 2 \\
\hline                                           
$\chi^{2}/ndf$                            &      0.67       \\ 
$f_{\rm cont}$                              & $(9.13\pm1.67\pm0.72)\times10^{4}$ GeV$^{n}$/pb \\
$n$                                                       & $5.33\pm0.13\pm0.02$              \\
$\rm{BF}(\psi(4040)\rightarrow\pi^+\pi^-\pi^+\pi^-\pi^0)$ & $(2.41\pm0.05\pm0.79)\%$          \\
$\phi$                                                    & $(267.70\pm0.44\pm9.53)^{\circ}$  \\
\hline
\end{tabular}
}
\label{table:fit_cont_4040}
\end{table}
\begin{figure}[htbp]
  \centering
  \includegraphics[width=0.45\textwidth]{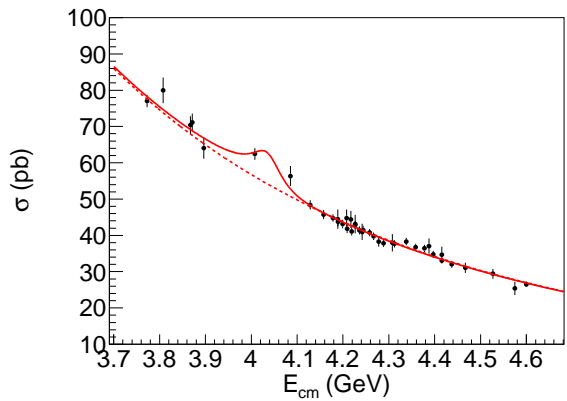}
  \caption{\sf
Fit to dressed cross sections for $e^+e^-\rightarrow\pi^+\pi^-\pi^+\pi^-\pi^0$ considering contribution from continuum process, $\psi(4040)$ decay and interference between them. Points with error bars show the measured dressed cross sections. The red curve shows the fit result. The dashed red curve shows the contribution from continuum process. 
}
  \label{fig:cs_fit_results_1stru}
\end{figure}
Compared to the previous fit result listed in Table~\ref{table:fit_cont}, the number of free parameters is increased by 2 (${\rm BF}(\psi(4040)\rightarrow\pi^+\pi^-\pi^+\pi^-\pi^0)$ and $\phi$) and the value of the $\chi^{2}$ is reduced by 16.0, which corresponds to a statistical significance of $3.6\sigma$.
\subsubsection{Upper limits of $Y(4230)$ decays}
Furthermore, we search for decays of the charmonium-like resonance $Y(4230)$ into those same final states, and corresponding upper limits are provided since no clear signal is observed. The expected dressed cross sections are constructed as:
\begin{small}
\begin{equation}
\sigma^{\rm expected}_{\rm{cont}(+\psi(4040))+Y(4230)}=|A_{\rm cont}(+A_{\psi(4040)}e^{i\phi})+A_{Y(4230)}e^{i\phi'}|^{2},
\end{equation}
\end{small}
where $\phi$ and $\phi'$ are the relative phases. $A_{\psi(4040)}$ and are Breit-Wigners for $\psi(4040)$ and $Y(4230)$, respectively.
The resulting likelihood distribution as a function of the $Y(4230)$ yield is used to get the upper limits at  
$90\%$ confidence level for ${\rm BF}(Y(4230)\rightarrow\rm{Light~Hadrons})$. The integral from zero to the upper limit contains $90\%$ of the area of the likelihood distribution. 
The original likelihood distributions are convolved with a Gaussian function, whose width is the systematic uncertainty of the cross section.
To estimate the systematic uncertainties in the upper limits of $Y(4230)$ decays, we have used values of $M_{Y(4230)}$ and $\Gamma^{\rm total}_{Y(4230)}$ from different experimental measurements~\cite{Y4220_Jpsipipi,Y4220_omegachic,Y4220_hcpipi,Y4220_DDstarpi}, and the largest resulting upper limits of ${\rm BF}(Y(4230)\rightarrow\rm{Light~Hadrons})$ are given in Table~\ref{table:upper}.
\begin{table}[htbp]\small
\centering
\caption{Summary of upper limits of $Y(4230)$ decays at $90\%$ confidence level.}
\resizebox{0.45\textwidth}{!}{
\begin{tabular}{|c|c|c|} \hline
Final state  &  $\Gamma^{ee}_{Y(4230)}\times{\rm BF}(Y(4230)\rightarrow\rm{Final~State})$(eV) \\
\hline                                           
$K^+K^-\pi^+\pi^-$         & $<19.7035$ \\
$K^+K^-K^+K^-$             & $< 3.8400$ \\
$\pi^+\pi^-\pi^+\pi^-$     & $<31.9727$ \\
$p\bar{p}\pi^+\pi^-$       & $< 7.2372$ \\
$K^+K^-\pi^+\pi^-\pi^0$    & $<43.3625$ \\
$K^+K^-K^+K^-\pi^0$        & $< 2.1206$ \\
$\pi^+\pi^-\pi^+\pi^-\pi^0$& $<16.0925$ \\
$p\bar{p}\pi^+\pi^-\pi^0$  & $<14.8893$ \\
\hline
\end{tabular}
}
\label{table:upper}
\end{table}
\section{Systematic Uncertainty}
\label{sec:sys}
The systematic uncertainties on the cross section measurements mainly come from the uncertainties in tracking, $\pi^{0}$ reconstruction, luminosity, fit of $\chi^2$ distribution, $E_{\rm{EMC}}/p$ cut, $\cos\theta_{\pi^+\pi^-}$ cut, and efficiency determination.\par
The uncertainty of the tracking efficiency is $1.0\%$~\cite{tracking} per track. The uncertainty of $\pi^{0}$ reconstruction is $2.0\%$~\cite{pi0}. The luminosity is measured using Bhabha events, with an uncertainty of $1.0\%$\cite{Lumxyz}.
To determine the systematic uncertainty due to PID requirements, we selected control samples of $K^+K^-\pi^+\pi^-$ and $K^+K^-\pi^+\pi^-\pi^0$ from data and MC. The MC sample is the PHSP MC reweighted accroding to the amplitude analysis results. The nominal PID requirements are replaced with a tighter cut: for kaon (pion) selection, it is required the probability to be kaon (pion) is greater than the probability to be pion (kaon). 
By studying the change of efficiencies, we determine the systematic uncertainty due to PID requirements for $K^+K^-\pi^+\pi^-$ and $K^+K^-\pi^+\pi^-\pi^0$ final states are both $0.5\%$.\par
\begin{table*}[htbp]
\centering
\caption{Summary of systematic uncertainties of the cross sections (in units of $\%$). The uncorrelated systematic uncertainty is marked by $^{*}$.}
\resizebox{\textwidth}{!}{
\begin{tabular}{|c|c|c|c|c|c|c|c|c|} \hline
Source  & $K^+K^-\pi^+\pi^-$ & $K^+K^-K^+K^-$ & $\pi^+\pi^-\pi^+\pi^-$ & $p\bar{p}\pi^+\pi^-$ & $K^+K^-\pi^+\pi^-\pi^0$ & $K^+K^-K^+K^-\pi^0$ & $\pi^+\pi^-\pi^+\pi^-\pi^0$ & $p\bar{p}\pi^+\pi^-\pi^0$ \\
\hline                                           
Tracking                            & $4.0$& $4.0$ & $4.0$                  & $4.0$ & $4.0$ & $4.0$ & $4.0$ & $4.0$\\ 
$\pi^{0}$ reconstruction            &   ---  &   ---   &   ---                    &  ---    & $2.0$ & $2.0$ & $2.0$ & $2.0$\\
Luminosity                          & $1.0$& $1.0$ & $1.0$                  & $1.0$ & $1.0$ & $1.0$ & $1.0$ & $1.0$\\
$E_{\rm EMC}/p<0.8$                 & $0.1$& $0.1$ & $0.1$                  &  ---    &  ---    &  ---    & $0.1$ &  ---   \\
$\cos\theta_{\pi^+\pi^-}<0.9$       & $0.1$&   ---   & $0.3$                  & $0.2$ &  ---    &  ---    &  ---    &  ---   \\
PID                                 & $0.5$&   ---   &   ---                    &  ---    & $0.5$ &  ---    &  ---    &  ---   \\
                Fit                 & $1.2$& $1.5$ & $1.0$                  & $2.5$ & $3.0$ & $5.0$ & $1.5$ & $1.8$\\
$^{*}$Efficiency                    & $0.9$& $2.0$ & $1.3$                  & $1.5$ & $0.8$ & $2.6$ & $2.1$ & $0.8$\\
                                                                                
\hline                                                                            
Total & $4.4$& $4.8$ & $4.4$ & $5.1$ & $5.6$ & $7.3$ & $5.3$ & $5.0$\\
\hline
\end{tabular}
}
\label{table:sysError}
\end{table*}
The contamination rates from peaking backgrounds are less than $0.8\%$ at all energy points for all signal final states.
Those contamination rates are very low, so we neglect the systematic uncertainty from peaking background subtraction. The estimation of non-peaking background is from the fit to the $\chi^{2}$ distribution. To estimate the uncertainty from the fit to the $\chi^{2}$ distribution, we refit the $\chi^{2}$ distributions by varying the bin size, the fit range, the signal shape and the background shape. The shape of signal MC is determined through MC simulation. In this work, for every candidate event, we select the combination with the least $\chi^{2}$. But it may be the wrong combination, and hence affect the signal shape. So we replace the nominal signal $\chi^{2}$ shape with the one from MC truth information, compare with the nominal result, and determine the systematic uncertainty due to the selection of the wrong combination.  The helix parameters are used to describe the tracks. To reduce the systematic uncertainty, we corrected the helix parameters from MC samples. We change the helix correction factors by $\pm1\sigma$ to determine the systematic uncertainty due to the signal shape. The shape of non-peaking backgrounds are determined using inclusive MC samples. 
We replace the shape of the non-peaking backgrounds by an Argus function and refit the $\chi^{2}$ distributions. The difference on the cross sections is taken as the systematic uncertainty. By adding these values in quadrature, we assign the uncertainty associated with the fit, which is dominated by the background shape. The distributions of $E_{\rm{EMC}}/p$ and $\cos\theta_{\pi^+\pi^-}$ from MC sample have been corrected according to data, which corresponds to correction factors on the reconstruction efficiency. The errors of the correction factors are taken as the systematic uncertainties.\par 
When determining the efficiencies, we applied a AmpTools analysis using data and MC samples at 4.226 GeV, and fixed the model to get efficiencies at every energy point. To estimate the systematic uncertainty, we repeat the process using data and MC samples at 3.773 GeV, 4.008 GeV, 4.258 GeV, 4.358 GeV, 4.416 GeV and 4.600 GeV. The maximum difference in efficiency is taken as the systematic uncertainty. The uncertainty from the efficiency is treated as an uncorrelated uncertainty, which is taken into consideration in the analysis of cross sections.\par 
The values of $\kappa$ are related to input cross sections, which are obtained from fits. We randomly change the fit parameters according to the covariance matrix from the fit. $\kappa$ is recalculated according to different cross section parameters. The change in $\kappa$ is less than $0.025\%$ for all channels, so the uncertainty of $\kappa$ is negligible.\par
The systematic uncertainties of the fit parameters, as listed in Table~\ref{table:fit_cont} and Table~\ref{table:fit_cont_4040}, originate from the uncertainty of the c.m.~energies, the uncertainty of cross sections, and the uncertainty of $\psi(4040)$ resonance parameters.
The systematic uncertainty originating from c.m.~energies is estimated by smearing c.m.~energies with a standard deviation of 0.8 MeV~\cite{cms}. Then we take the maximum difference of the parameters as the systematic uncertainty.
Similarly, the systematic uncertainties due to cross sections and $\psi(4040)$ resonance parameters are estimated by re-obtaining the fit parameters after changing the value of the cross sections and $\psi(4040)$ resonance parameters by $\pm1\sigma$, respectively. Finally, we assign the total systematic uncertainty by adding these values in quadrature.\par 
\section{Summary}
\label{sec:summary}
The dressed cross sections for the processes $e^+e^-\rightarrow K^+K^-\pi^+\pi^-(\pi^0)$, $K^+K^-K^+K^-(\pi^0)$, 
	$\pi^+\pi^-\pi^+\pi^-(\pi^0)$, $p\bar{p}\pi^+\pi^-(\pi^0)$ are obtained with data samples collected at 40 energy points from 3.773 to 4.600 GeV. Those
cross sections depend on c.m.~energy according to $1/E_{\mathrm cm}^n$ and we determine $n$ for each process.
We find evidence of the $\psi(4040)$ decays to $\pi^+\pi^-\pi^+\pi^-\pi^0$ with a statistical 
significance of $3.6\sigma$. No obvious $Y(4230)$ signals are observed, so we provide upper limits for $Y(4230)$ decays into those final states at the $90\%$ confidence level.

\section*{Acknowledgments}
The BESIII collaboration thanks the staff of BEPCII and the IHEP computing center for their strong support. This work is supported in part by National Key Basic Research Program of China under Contract No. 2015CB856700; National Natural Science Foundation of China (NSFC) under Contracts Nos. 11521505, 11625523, 11635010, 11735014, 11822506, 11835012, 11935015, 11935016, 11935018, 11961141012; the Chinese Academy of Sciences (CAS) Large-Scale Scientific Facility Program; Joint Large-Scale Scientific Facility Funds of the NSFC and CAS under Contracts Nos. U1732263, U1832207; CAS Key Research Program of Frontier Sciences under Contracts Nos. QYZDJ-SSW-SLH003, QYZDJ-SSW-SLH040; 100 Talents Program of CAS; National 1000 Talents Program of China; INPAC and Shanghai Key Laboratory for Particle Physics and Cosmology; ERC under Contract No. 758462; German Research Foundation DFG under Contracts Nos. Collaborative Research Center CRC 1044, FOR 2359; Istituto Nazionale di Fisica Nucleare, Italy; Ministry of Development of Turkey under Contract No. DPT2006K-120470; National Science and Technology fund; STFC (United Kingdom); The Knut and Alice Wallenberg Foundation (Sweden) under Contract No. 2016.0157; The Royal Society, UK under Contracts Nos. DH140054, DH160214; The Swedish Research Council; U. S. Department of Energy under Contracts Nos. DE-FG02-05ER41374, DE-SC-0012069.\par

\end{document}